\documentclass[11pt,preprint]{aastex}

\newcommand{\lllambda}{\ensuremath{\lambda L_{\lambda}}}

\newcommand{\fnu}{\ensuremath{f_{\nu}}}

\newcommand{\lya}{\ensuremath{{\rm Ly}{\alpha}}}

\newcommand{\ha}{\ensuremath{{\rm H}{\alpha}}}

\newcommand{\hi}{H\,{\sc i}}

\newcommand{\oiii}{[O\,{\sc iii}]}
\newcommand{\oiiil}{[O\,{\sc iii}]\,$\lambda$5007}

\newcommand{\feii}{Fe\,{\sc ii}}

\newcommand{\vla}{{\it VLA}}

\newcommand{\chandra}{{\it Chandra}}
\newcommand{\spitzer}{{\it Spitzer}}
\newcommand{\xmm}{{\it XMM}}

\newcommand{\tmass}{{\it 2MASS}}

\newcommand{\rosat}{{\it ROSAT}}
\newcommand{\hst}{{\it HST}}
\newcommand{\fuse}{{\it FUSE}}

\newcommand{\iso}{{\it ISO}}
\newcommand{\galex}{{\it GALEX}}
\newcommand{\kms}{\ensuremath{{\rm km~s}^{-1}}}
\newcommand{\kev}{\ensuremath{\mbox{keV}}}

\newcommand{\fluxl}{erg s$^{-1}$ cm$^{-2}$ \AA$^{-1}$}
\newcommand{\fluxhz}{erg s$^{-1}$ cm$^{-2}$ Hz$^{-1}$}

\shorttitle{Quasar SEDs}
\shortauthors{Shang et al.}

\begin{document}

\title{The Next Generation Atlas of Quasar Spectral Energy Distributions 
from Radio to X-rays}


\author{Zhaohui Shang,\altaffilmark{1,2}
Michael S. Brotherton,\altaffilmark{2}
Beverley J. Wills,\altaffilmark{3}
D. Wills,\altaffilmark{3}
\\
Sabrina L. Cales,\altaffilmark{2}
Daniel A. Dale,\altaffilmark{2}
Richard F. Green,\altaffilmark{4}
Jessie C. Runnoe,\altaffilmark{2}
\\
Rodrigo S. Nemmen,\altaffilmark{5}
Sarah C. Gallagher,\altaffilmark{6}
Rajib Ganguly,\altaffilmark{7}
Dean C. Hines,\altaffilmark{8}
\\
Benjamin J. Kelly,\altaffilmark{2}
Gerard A. Kriss,\altaffilmark{9,10}
Jun Li,\altaffilmark{1}
Baitian Tang,\altaffilmark{11,12}
Yanxia Xie,\altaffilmark{13,1}
}

\altaffiltext{1}{Department of Physics, Tianjin Normal University,
Tianjin 300387, China. zshang@gmail.com}
\altaffiltext{2}{Department of Physics and Astronomy, University of Wyoming,
Laramie, WY 82071, USA}
\altaffiltext{3}{Department of Astronomy, University of Texas at
Austin, 1 University Station, C1400, Austin, TX 78712.}
\altaffiltext{4}{Large Binocular Telescope Observatory, University of
Arizona, 933 N. Cherry Ave., Tucson, AZ 85721, USA}
\altaffiltext{5}{NASA/Goddard Space Flight Center, 
Greenbelt, Maryland 20771 USA}
\altaffiltext{6}{Department of Physics \& Astronomy, The University of
Western Ontario, London, ON N6A 3K7, Canada}
\altaffiltext{7}{Department of Computer Science, Engineering, \&
Physics, University of Michigan-Flint, 213 Murchie Science Building,
303 Kearsley Street, Flint, MI 48502, USA}
\altaffiltext{8}{Space Science Institute, 4750 Walnut Street, Suite
205 Boulder, CO 80301, USA}
\altaffiltext{9}{Space Telescope Science Institute, 3700 San Martin
Drive, Baltimore, MD 21218, USA}
\altaffiltext{10}{Department of Physics and Astronomy, 
The Johns Hopkins University, Baltimore, MD 21218}
\altaffiltext{11}{Department of Physics, 1245 Webster Hall, Washington
State University, Pullman, WA, 99164-2814 USA}
\altaffiltext{12}{Department of Astronomy, Nanjing University,
Nanjing, 210093, China}
\altaffiltext{13}{Shanghai Astronomical Observatory, 
Chinese Academy of Science, 80 Nandan Road, Shanghai 200030, China}

\begin{abstract}

We have produced the next generation of quasar spectral energy
distributions (SEDs),  essentially updating the work of 
Elvis et al.\ (1994) by using high-quality data obtained with 
several space and ground-based telescopes, including
NASA's Great Observatories.
We present an atlas of
SEDs of 85 optically bright, non-blazar quasars over the electromagnetic
spectrum from radio to X-rays.  The heterogeneous sample includes 27 
radio-quiet and 58 radio-loud quasars.
Most objects have quasi-simultaneous ultraviolet-optical spectroscopic data, 
supplemented with some far-ultraviolet spectra, and more than half also have
\spitzer\ mid-infrared IRS spectra.  The X-ray spectral parameters are
collected from the literature where available.  The radio, far-infrared,
and near-infrared photometric data are also obtained from either
the literature or new observations.  We construct composite spectral
energy distributions for radio-loud and radio-quiet objects and compare
these to those of Elvis et al., finding that ours have similar overall shapes, 
but our improved spectral resolution reveals more detailed features, 
especially in the mid and near-infrared.  


\end{abstract}



\section{INTRODUCTION}


The supermassive black holes powering quasars (or active galactic
nuclei (AGN)) do not themselves shine.  It is the heated material
surrounding the black holes that emits the radiation signatures of
quasars.  These signatures include broad emission lines characteristic
of high-velocity gas moving at thousands of kilometers per second, 
and extremely high continuum luminosity in
excess of that of entire galaxies.  Continuum emission is seen
in all parts of the electromagnetic spectrum from the highest energies
(gamma rays and X-rays) to the lowest (radio waves).  The power
emitted is similarly high, within an order of magnitude or so,
over much of this range, although there can be significant variation from
quasar to quasar. 

There are no single states of matter or single processes capable of 
reproducing the spectral energy distribution (SED) of a quasar.  
A combination of both thermal and non-thermal processes has been
invoked to explain, at different parts of the SED, the emission from gas 
in a variety of states, at a variety of temperatures, at a variety of 
distances, and experiencing a variety of environments.
Quasars seem to have many different components that are expressed in
different parts of the electromagnetic spectrum.



So what are the components of the quasar?  The central supermassive black
hole is the ultimate engine, allowing the liberation of gravitational
potential energy.  The primary
source of electromagnetic emission is likely an accretion disk 
formed of hot gas spiraling
into the black hole and shining in the optical through ultraviolet
(UV).  A hot atmosphere upscatters the disk photons to X-ray energies.
A jet shooting out from the inner accretion disk emits synchrotron
radiation, dominating radio emission and sometimes higher energies.
An obscuring torus of relatively cool gas and dust, heated by photons
from the accretion disk, thermally radiates in the near-infrared (NIR) and
mid-infrared (MIR).  The far-infrared (FIR) part of the SED comes from
cooler dust, perhaps distributed throughout the host galaxy, that may be heated by stars rather than the quasar
itself.  Other regions in and among these continuum-emitting parts are
responsible for the prominent emission lines present in quasar
spectra. The details of all these parts, mechanisms, and their
relationships are not yet completely understood, 
because multi-wavelength
data to further our understanding have been difficult to gather.

Multiwavelength astronomy is challenging.  No single telescope
can observe at all wavelengths.  Many parts of the electromagnetic 
spectrum cannot penetrate the Earth's atmosphere and require space-based
observations.  Quasars are variable and this means that simultaneous
or nearly simultaneous observations are desirable, at least in some 
parts of the spectrum.  Different technologies have different sensitivity 
levels and what may be an easily observed target at one wavelength
may be difficult to detect at another.  

There were a number of pioneering works on quasar SEDs  in the 1980s 
\citep[e.g.,][]{EdeMal86,Ward87,Kriss88,Sande89,SunMal89}.  In the 1990s, 
\citet[][hereafter E94]{Elvis94}
established the first large, high-quality atlas of quasar SEDs.
The timing of their work was predicated on the launch of several space-based
telescopes, notably IRAS in the mid to far infrared, IUE in the UV,
and Einstein in the X-rays, that for the first time provided
observations of large numbers of quasars in these wavebands.  They
established the differences between the SEDs of radio-loud (RL) and
radio-quiet (RQ) quasars.  They also characterized the variance of quasar
SEDs, which is rather substantial, and explored the problem of
bolometric corrections. 

Since the work of E94, there have been several
significant investigations of quasar SEDs
\citep[e.g.,][]{Kuras03,RisElv04}.  \citet{Richa06} is
the largest that covers the entire electromagnetic spectrum,
including data from the Sloan Digital Sky Survey and \spitzer,
supplemented by near-IR, GALEX ultraviolet, VLA radio, and ROSAT X-ray
data, where available.
One of their key findings was again the wide range of SED shapes, and
how assuming a mean SED can potentially lead to errors in bolometric
luminosities as high as 50\%.

Other investigations of SEDs have focused on subclasses, like broad
absorption line (BAL) quasars \citep[e.g.,][]{Galla07}, 2MASS red
quasars \citep{Kuras09}, hard X-ray selected quasars \citep{Polle00,
Polle07}, or on individual objects \citep[e.g.,][]{Zheng01}.  More
detailed SED work has been done on limited portions of the entire
electromagnetic spectrum, such as the optical-X-ray
\citep[e.g.,][]{Laor97}, 
the FIR to optical \citep[e.g.,][]{Netze07}, or optical/UV to X-ray
\citep{Grupe10}.


There has also been work looking for relationships among more detailed
spectral features, like emission lines and SEDs.  \citet{Wilke99},
examining 41 quasars with SED information, for instance, 
found a variety of Baldwin effects (Baldwin 1977), anticorrelations between
emission-line equivalent width and UV luminosity, as well as some
correlations between properties of Fe II and C IV emission lines
and the ratio of the optical to X-ray luminosity.  \citet{Schwe06} and 
\citet{Netze07}
studied Palomar-Green quasars with FIR to optical data and confirmed
that most FIR radiation is due to star-forming activity.  However,
\citet{Netze07} also argue, based on a correlation between
$L(5100\mbox{\AA})$ and $L(60\micron)$,
an alternative view that a large fraction
of FIR radiation could result from direct AGN heating.

Ideally what one would like in studying quasars is a complete inventory
over all time and all directions of all photons emitted. The best we can
do now is to obtain spectrophotometric snapshots, close in time, of some
spectral regions, supplemented by photometry in other accessible wavebands,
from our particular line of sight toward a quasar.
The technology has improved since the 1990s, with spacecraft such as 
HST, Chandra and XMM, and Spitzer replacing IUE, Einstein and ROSAT, 
and IRAS, respectively.

This atlas is meant to update the last decade's work with a modern set 
of quasar SEDs using the next generation of telescopes and instruments.
Whenever possible we have striven to use high-quality spectrophotometry
in addition to photometry, which will enable investigations like those
of Wilkes et al. (1999).  One hope is that spectral features will be
found that correlate with SEDs and allow their shapes to be 
determined without the need for complete multi-wavelength observations.
Any correlations found should provide deeper insight into quasars.
We plan additional papers based on this data set performing 
various types of investigations, as well as addressing both observational
and theoretical aspects of bolometric corrections.

In the next section (\textsection\ref{sec:sample}), we describe our sample, which is composed of three
subsamples that have excellent quasi-simultaneous optical through UV
or far-UV (FUV) spectrophotometry serving as a starting point for
SED construction.  We then describe the sample properties.  Subsequent
sections describe the data (\S\ref{sec:data}), starting with about
100~MHz radio and moving to
higher energies to about 10\,\kev\ X-ray.  We then discuss corrections
to the SEDs, such as Galactic dereddening and host galaxy removal
(\S\ref{sec:correction}). 
In \S\ref{sec:sed}, we present the SEDs for our quasars individually, then as
composites for radio-loud and radio-quiet subsamples, which are known
to differ significantly.
We then discuss the properties of the SEDs, and compare our composites
with those of E94.  We finish this paper with a summary of these
results, future plans, and some concluding remarks (\S\ref{sec:summary}).  


\section{SAMPLE\label{sec:sample}}

In the past, in order to study the optical and ultraviolet properties
of quasars, we embarked on several programs of obtaining
quasi-simultaneous spectrophotometry utilizing various ground and
space-based telescopes.  As this is a challenging endeavor, and the
data we obtained were of high quality, we began the construction of
SEDs with these samples, adding data at longer and shorter
wavelengths.  For convenience, we refer to our three subsamples by the
abbreviations PGX, FUSE-HST, and RLQ.  These subsamples are described
below, and Table~\ref{tb:sample} provides our combined SED sample of
85 objects and their basic properties.  

Several objects in the FUSE-HST subsample are also in the other two
subsamples and there may be repeated observations in UV and/or
optical.  We choose to analyze the data with the FUSE-HST subsample
because of their higher quality.

\subsection{PGX}

The ``PGX'' sample consists of 22 of 23 Palomar-Green quasars in the complete sample
selected by \citet{L94,Laor97} from the Bright Quasar Survey
\citep[BQS;][]{SchGre83}. Interested in observing the soft-X-ray
regime using bright quasars, \citet{L94,Laor97} started with the
ultraviolet-excess selected BQS and added the restrictions that
$z\leq0.4$ and Galactic \hi\ column density $N_{\rm HI} < 1.9\times
10^{20} {\rm cm} ^{-2}$.  We obtained low-resolution ultraviolet
spectra with \hst\ and conducted quasi-simultaneous ground-based
observations at McDonald Observatory, usually within a month.
\citet{Shang07} provide additional details and constructed the
UV-optical SEDs for this sample.


\subsection{FUSE-HST}

The ``FUSE-HST'' sample originates with the \fuse\ AGN program
\citep{Kriss00}, which surveyed more than 100 of the UV-brightest
AGNs.  About 20 of these were also observed in an \hst\ spectral
snapshot survey with sufficient signal-to-noise ratio 
during 1999--2000.  The \fuse\ observations were
scheduled as close in time as possible with the \hst\ snapshot
observations, and ground-based optical spectra were also obtained
during the same period at Kitt Peak National Observatory (KPNO).  We
exclude a few objects because of the lack of an optical spectrum
(NGC~3783, low declination), very strong host galaxy contamination
(NGC~3516), or strong variability (NGC~5548, also no simultaneous
\hst\ spectrum).  Our final FUSE-HST sample includes 17 objects with
quasi-simultaneous spectrophotometry extending to the FUV and covering
rest wavelength from 900--9000\,\AA.  This is a heterogeneous sample
with low redshift ($z<0.5$).  \citet{Shang05} provides additional
details.

\subsection{RLQ}

The ``RLQ'' sample originates with an early \hst\ program to observe a
large sample of radio-loud quasars selected to have a small range in
extended radio luminosity, a property thought to be isotropic.
Limitations of \hst\ discovered after launch required adjustments to
the sample and brighter objects were substituted for fainter ones. Over the
course of four cycles, \hst\ targeted nearly 50 quasars.
Quasi-simultaneous optical spectrophotometry was obtained at several
observatories, primarily McDonald Observatory and KPNO.
\citet{Wills95} and \citet{Netze95} provide additional details of the
sample.  A number of the radio-core dominant quasars are
blazars, with optically violent variability due to synchrotron
emission from a beamed jet.  We have excluded these blazars from the
sample based on rapid optical variability as we regard this component
as a major complication in determining intrinsic and uniform SEDs 
for comparison.

\subsection{Sample Properties \label{sec:property}}

\begin{figure}
\includegraphics[angle=270,scale=0.6]{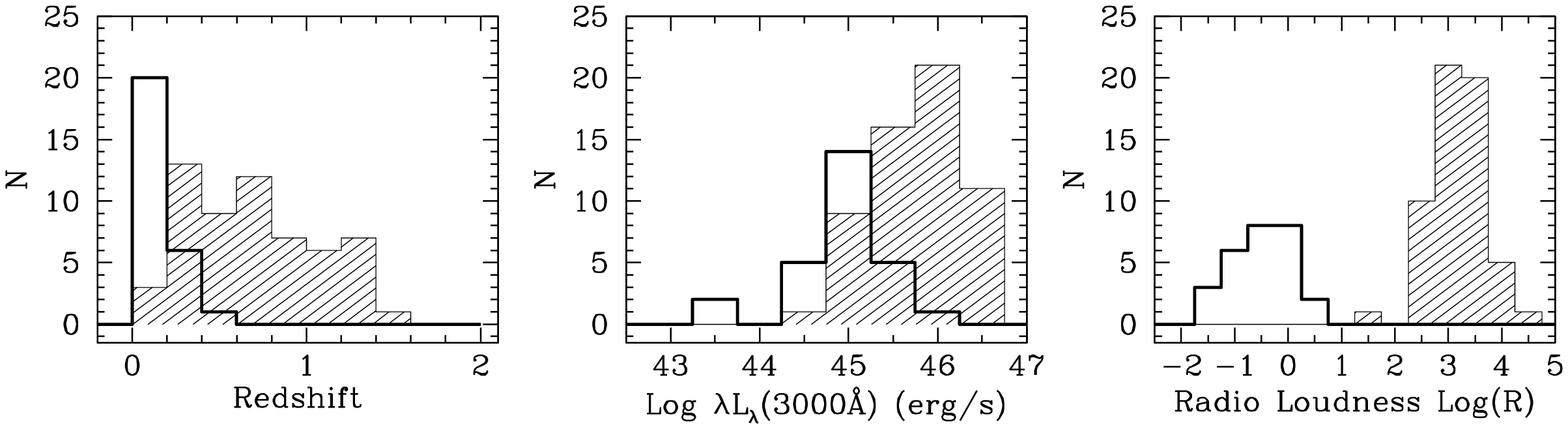}
\caption{Distributions of sample properties for radio-loud and
radio-quiet sub-samples.
Shaded bins are for radio-loud objects.
}
\label{fg:property}
\label{fg:hist}
\end{figure}

In order to summarize the properties of the combined sample,
we have plotted some histograms (Fig.~\ref{fg:hist}).  We have
distinguished radio-loud and radio-quiet quasars using radio-loudness
calculated with our data.  We have also measured the redshift and
3000\,\AA\ rest-frame continuum luminosity of the sample.  Details are
provided in \S\ref{sec:calculation}. 

Of the 85 objects in the final sample, there are 27 RQ and 58 RL
quasars.  All RQ quasars are from either PGX or FUSE-HST subsamples,
having redshift less than 0.5.  Most RL quasars are from the RLQ
subsample and more than half of the RL quasars have redshift larger than
0.5.  

Both RL and RQ samples span about 2 orders of magnitude in 
luminosity.
The RL sample has an average luminosity about 6 times
higher.  These properties reflect the original selections of the
subsamples.

We emphasize that as a whole, this sample is representative of
UV/optical-bright quasars, both radio-loud and radio-quiet, but is not
statistically complete or well-matched.  Particular subsamples may be
appropriate for general statistical studies and comparisons only if
care is taken in their selection.

\section{DATA\label{sec:data}}

\begin{figure}
\includegraphics[angle=270,scale=0.5]{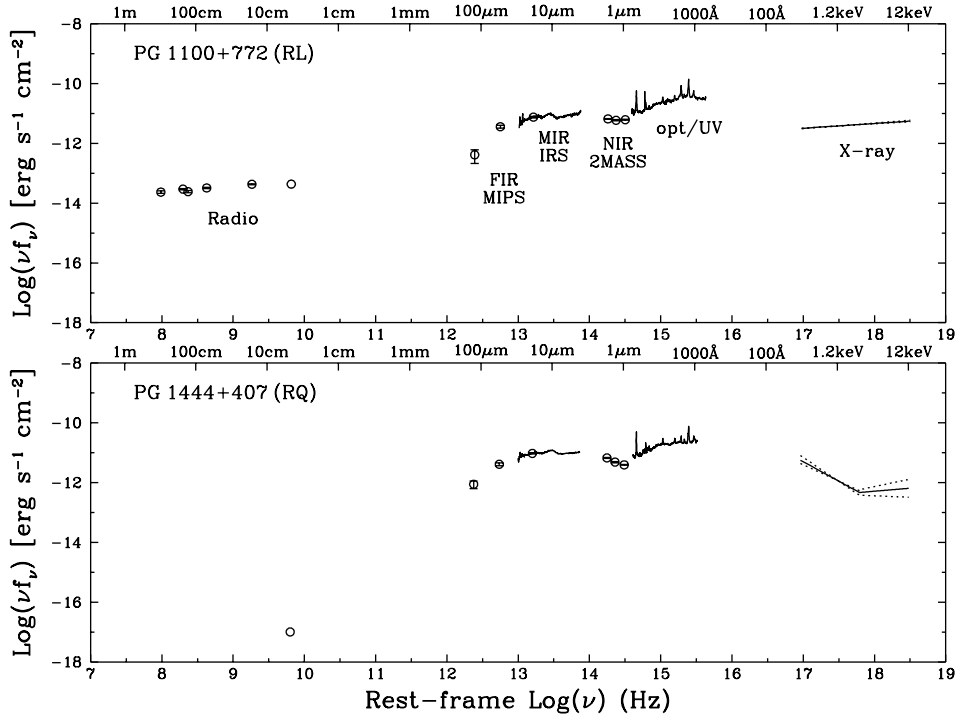}
\caption{Examples of our collected data for two objects.  Also marked
are the wavebands and some instruments used to obtain the data.
}
\label{fg:plot2}
\notetoeditor{Please display it for the full width of the page}
\end{figure}

We collected both photometry and spectroscopy data for this work.
Many data were obtained with space telescopes, including \hst, \fuse,
\spitzer, \chandra, and \xmm.  This ensures the unprecedented quality
of the SEDs.  Figure~\ref{fg:plot2} shows two examples of our objects,
marked with wavebands and instruments used to obtain the data.
These will be discussed in detail in the following sections.

Most of the data were obtained between 1991 and 2007 except for some
archival radio data which were obtained much earlier (see
Table~\ref{tb:radiosurvey} for details).
However, the problem of AGN
intrinsic variability can be neglected statistically for the SED work,
especially with regard to some portions (e.g., infrared) 
which have only very long timescale
variation.
Moreover, the FUV-UV-optical spectra of our sample were mostly
obtained quasi-simultaneously, within weeks, 
specifically minimizing this problem.

Before we combine data from different bands to construct the SEDs, we
applied two corrections: host galaxy correction in the near-IR
(\S\ref{sec:host}) and Galactic reddening in the FUV-UV-optical
spectra (\S\ref{sec:redden}).  We also derived some sample properties
from the data set (\S\ref{sec:calculation}).

\subsection{Radio\label{sec:radio}}

We collect radio data for the sample from archives of high-quality
surveys and some data from the literature.  Table~\ref{tb:radiosurvey}
lists all the references.  We choose the frequency range from 74~MHz
to about 15~GHz, where most objects have observations.  We include
surveys of similar frequencies (e.g., 325 and 365~MHz; 4850, 4890,
4990, and 5000~MHz) in order to maximize the number of objects with
available observations in a similar frequency range.

The total fluxes at each frequency are listed in Table~\ref{tb:radio}
and used in the SED construction and analysis.  When a survey or an
observation resolves the core and lobes, we make sure to get the total
flux by including all the components.  In doing so, we check the
positions of each component and take advantage of the higher
resolution map (5\arcsec) of FIRST survey for comparison.  
At 1400~MHz, we use the total flux density from NVSS in preference to
that from FIRST for all but three objects (IRAS~F07546+3928, Mrk~506,
and PG~1115+407) that are not included in the NVSS catalog.  NVSS has
a spatial resolution comparable to other major radio surveys we use,
and generally includes all the radio flux, while FIRST resolves many
sources into multiple regions of emission.  We did check the FIRST
images and collected the measurements for each object, which, when the
pieces are summed, give results consistent with those from NVSS.

Most radio-quiet objects are not detected in the radio surveys, which can
provide only an upper limit.  We therefore search the literature to
obtain at least one detection from individual studies.


\subsection{Far-IR\label{sec:mips}}

Far-infrared photometry at 24, 70, and 160\,\micron\ from 
Multiband Imaging Photometer for \spitzer\ \citep[MIPS,][]{Rieke04}
is available for 50 objects in our sample (Table~\ref{tb:mips}).
In addition to archive data, we obtained new data explicitly for this
study.  All these observations were made with MIPS photometry mode.

We perform aperture photometry on each object (point source) in all
three bands and apply corresponding aperture corrections based on the
aperture radius and sky annulus sizes listed in the MIPS Instrument
Handbook.  Although the MIPS handbook quotes a 10\% flux calibration
uncertainty for bright sources, the photometry uncertainties can be up
to 16\% for 24\micron, 40\% for 70\micron, and 60\% for 160\micron\
for the faint objects in our sample.  Moreover, several objects are
not detected at the 160\,\micron\ and we estimated a 3$\sigma$ upper
limit for them, where $\sigma$ is the standard deviation of the sky
background around the source position.

\subsection{Mid-IR\label{sec:mir} \label{sec:irs}}

The mid-infrared spectroscopy from the \spitzer\ Infrared Spectrograph
\citep[IRS,][]{Houck04,Werne04}
were obtained for 46 objects from both archival observations and new
observations made for this study.  We use data from 4 low-resolution
modules covering observed wavelengths from 5--40\micron.

\begin{figure}
\plotone{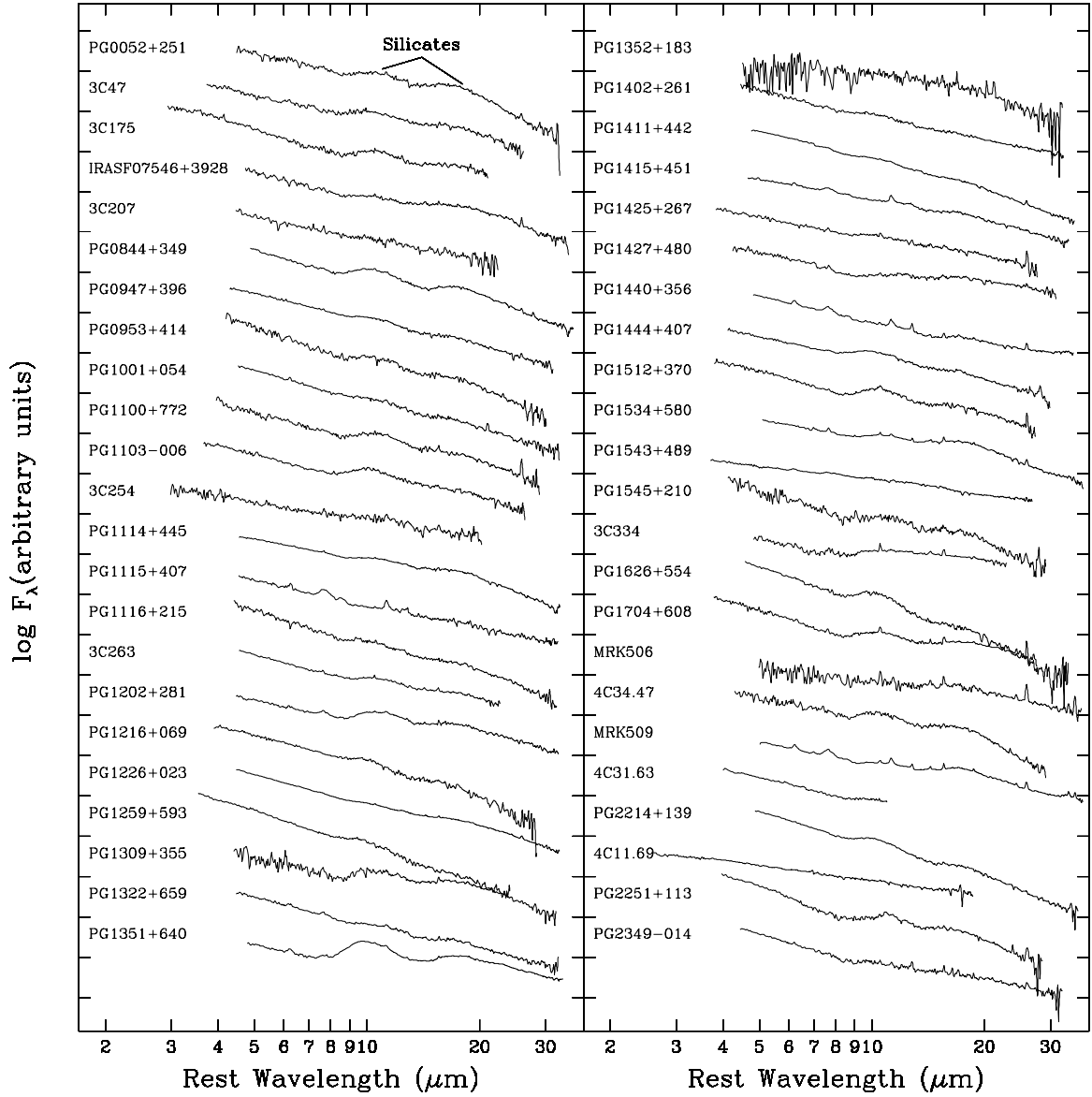}
\caption{Available Mid-Infrared spectra from \spitzer\ IRS for our sample.
The bumps around 10 and 18 \micron\ are silicate features.
}
\label{fg:mir}
\label{fg:irs}
\notetoeditor{Please display it for the full width of the page}
\end{figure}

Since all our objects are essentially point sources, we obtained the
spectra from standard post-basic calibrated data (PBCD) 
products.  Before we combined individual
segments from different wavebands, we removed flagged data points and
obvious spurious points at the edges of detectors.  The final spectra
are shown in Figure~\ref{fg:mir}, covering rest wavelength from about
$\sim3-35$\micron\ for the redshifts of our sample.  The spectrophotometric
calibration uncertainty is within 15\%, and this is also verified with
our 24\micron\ photometry of MIPS.

MIR spectra show clear broad silicate emission features around 10 and
18\micron, and narrow emission lines such as  
[\ion{S}{4}]$\lambda 10.5\micron$, 
[\ion{Ne}{5}]$\lambda 14.3\micron$, 
[\ion{Ne}{3}]$\lambda 15.6\micron$, 
and
[\ion{O}{4}]$\lambda 25.9\micron$.
These features have been investigated in detail in previous studies 
\citep[e.g.,][]{Hao05,Weedm05,Dale09,GouAle09,Diamo09,Perei10,Tomma10}, 
and have been kept in our analyses.

\subsection{Near-IR\label{sec:nir} \label{sec:2mass}}

We rely on 2MASS photometry in the near-infrared \citep{Skrut06},
supplemented by our own observations with NASA's
IRTF\footnote{The Infrared Telescope Facility is
operated by the University of Hawaii under Cooperative Agreement no.
NNX-08AE38A with the National Aeronautics and Space Administration,
Science Mission Directorate, Planetary Astronomy Program.}
and the Hubble
Space Telescope.  The 2MASS point-source catalog has 79 members of our
sample.  Table~\ref{tb:2massmag} provides magnitudes derived from
both point-source profile fitting
and aperture photometry from 2MASS, along with a
flag (ext) that indicates the source is also listed in the 2MASS
extended source catalog and then the aperture magnitudes are obtained from
there instead.  For objects only in the 2MASS point source catalog, the
apertures are 4\arcsec\ in radius, while for objects in the
extended source catalog the apertures are 14\arcsec\ in radius.

In the absence of additional information, we use the 2MASS point
source profile fitting magnitudes as the AGN magnitudes, but in many
cases we can do better than this.  Using host galaxy measurements, we
estimated and subtracted the host galaxy contribution to obtain AGN
magnitudes as described in \S~\ref{sec:host}.


\subsection{Near-UV -- Optical\label{sec:uvopt}}

We have UV-optical spectrophotometry for all the objects from our
previous studies \citep{Wills95,Netze95,Shang05,Shang07}.  We follow the
general observing and data reduction procedures to obtain the spectra.
We give a brief summary here.

The optical spectra were obtained from ground-based telescopes in
long-slit mode.  They were re-analyzed in a consistent way for all the
three subsamples.  The host galaxy contribution was checked carefully
and removed as much as possible when extracting the spectra.  The host
contribution can only significantly affect the red part of the optical
spectra and we used different aperture sizes to verify that the host
galaxy contamination in the final spectra is undetectable.  For
several higher redshift objects in the RLQ sample, we also obtained
near-IR spectra from UKIRT to cover the important rest-frame \ha\
regions.

The near-UV spectra are from HST Faint Object Spectrograph (FOS) 
for the RLQ and PGX subsamples over
several cycles.  For the FUSE-HST sample, the spectra are from 
Space Telescope Imaging Spectrograph (STIS) Snap
programs \citep{Kriss00}.  Most of these UV spectra were obtained
quasi-simultaneously (within weeks) with our optical spectra to reduce
the uncertainty caused by their intrinsic variability.  The standard
flux calibration are very good and are usually consistent with our
optical data \citep{Shang05,Shang07}. 
The typical flux density
uncertainty is less than 5\%.




\subsection{Far UV\label{sec:fuv}}

While the far-ultraviolet (FUV) portion of the SED is relatively
narrow,
it is of 
great interest for  several reasons.  The turnover and energy peak of the 
optical-UV ``big blue bump'' is in or near the FUV.
This portion of the SED is the part we can observe in most quasars that 
is closest to the peak of the ionizing continuum.  The ionizing continuum 
powers emission lines that have been used to estimate black hole masses and 
probably also drives high-velocity outflows that interact with the 
environment. 

We provide FUV data when available.  High-resolution observed-frame
FUV spectra, from 905-1187\,\AA, is available from \fuse\
\citep{Moos00} for a fraction of our sample, primarily the FUSE-HST
subsample (17 objects).  \citet{Shang05} provide details about the
\fuse\ data of this subsample and cite additional related technical
material about \fuse.


There was also a \fuse\ program specifically targeting the PGX sample
but most of these objects turned out to be too faint for 
\fuse.  In addition to the FUSE-HST sample, the \fuse\
archive does have observations of several additional quasars in our
sample, good enough for SED purposes: 3C 263, PG 1116+215, PG
1216+069, PG 1402+261, PG 1415+451, PG 1440+356, PG 1626+554, and
these data are included in the same way as in \citet{Shang05}.
All 24 objects with \fuse\ data and their rest-frame wavelength
coverage in \fuse\ are listed in Table~\ref{tb:fuse}.

Our sample with \fuse\ data will be biased to lower luminosity, lower
redshift objects, typically bright Seyfert 1 galaxies.  Higher redshift
quasars will fortunately have the rest-frame FUV redshifted to longer
wavelengths observable with \hst, so this bias is not very significant.

\subsection{X-ray\label{sec:xray}}

The X-ray data are collected from \chandra, \xmm\ and \rosat\
sources reported in the literature.  
Because of their higher sensitivity and broader energy
coverage, we always choose \chandra\ and \xmm\ data when available;
otherwise, we resort to \rosat.  We have a total 71 objects with
X-ray information, 34 from \rosat\ data.

For individual X-ray studies of AGNs, the data are usually fitted with
different models and components to reveal detailed X-ray
properties.   Sometimes the models are very complicated, but for the
purpose of SED work, we focus on the overall shape of the energy
distribution in this region, therefore, we try to choose the simplest,
best fitting models.  This includes either a single power-law or a
broken power-law.   In addition to individual studies, we also
obtained the results of 3 objects from the Chandra Source Catalog
\citep[CSC,][]{Evans10}.

The spectral indices and flux densities from different studies have been
converted to a uniform system for consistency, 
$f_\nu = f_0 E^\alpha$, 
where $f_0$ is the flux density at 1\,\kev, in units of
\fluxhz, E in \kev, and $\alpha$ is the power-law (or broken
power-law) spectral index.  The results are listed in
Table~\ref{tb:xray} along with the references.

In order to obtain the SED in the X-ray domain,  we re-build the
power-law or broken power-law ``spectra'' using the spectral index
$\alpha$ and  $f_0$ within the instrument-related energy ranges in the
observed-frame.  A sampling of 0.1\,\kev\ is enough to show the X-ray
SED shape.  We also use the errors of $\alpha$ to estimate the
uncertainty of the X-ray SEDs.



\section{CORRECTIONS AND MEASUREMENTS\label{sec:correction}}

\subsection{Near-Infrared Host Galaxy Corrections \label{sec:host}}

Most of our objects are UV-optical bright quasars and the host galaxy
contamination to the AGN light is not large.  When possible, we have
tried to estimate the host galaxy contribution using photometry at
H-band, where, due to typical SED shapes of galaxies and quasars, the
fraction of host galaxy contribution may be maximized in contrast to
AGN light at the redshifts around 0.5 \citep[see Fig.~1
in][]{McLRie95} and is easier to detect.

For 33 sample members we have made our own observations, or used those
from the literature, in order to determine the H-band host galaxy
fractions.  These are given in Table~\ref{tb:2masshost}.

Our IRTF observations (5 objects), as well as those of \citet[][17
objects]{McLRie94a,McLRie94b} used ground-based telescopes, long
exposure times, and were obtained with seeing of 1-2 arcseconds.  In
general, infrared imaging is done by mosaicking together large numbers
of short exposure time images of the object on different positions on
the chip.  We reduced our data, generally 30 minute exposures on
target, using the DIMSUM task inside IRAF.  We determined our
host-galaxy fractions using a similar one-dimensional analysis
procedure to that of McLeod \& Rieke (1994a).  This includes fitting a
standard star observed just before or after the target to the
one-dimensional surface brightness profile.  Minimal and maximal
subtraction of standard star PSFs indicates an uncertainty in this
part of the procedure of just a few percent, which is small compared
with other systematic uncertainties.

We observed an additional 7 higher redshift sample members with NICMOS
on HST with H-band, and supplement these with 4 more similar
observations by McLeod \& McLeod (2001, two of these superseding
results from McLeod \& Rieke 1994b).  For the sharper and more regular
HST images, two-dimensional PSF fitting is possible.  Our observations
of the targets were for one orbit each and we also observed a standard
star for each target.  We used GALFIT \citep{Peng02} to fit the PSF to each image,
along with several galaxy models (e.g., exponential disks and
appropriately constrained Sersic profiles), and took the results from
the best fit.  The different methods provided host galaxy fractions
consistent to a few percent or better, again more than adequate for
SED work.

We choose the 2MASS aperture magnitude (Table~\ref{tb:2massmag}) as the
total magnitude of an AGN and its host.  It is straightforward to
correct the H-band host galaxy contamination once we have the measured
host fraction in H-band.  To correct J and K band magnitudes for host
galaxy contamination, we subtracted an appropriately scaled and
redshifted elliptical galaxy template (NGC 584 from Dale et al. 2007)
from the 2MASS aperture photometry.  

Table~\ref{tb:2masscorr} gives the final AGN J, H, and K magnitudes
used for the SEDs.  For the 33 objects with detailed host galaxy
corrections, the corrected magnitudes are listed.  For the rest of the
objects, the 2MASS PSF magnitudes (profile fitting magnitudes) are
used.

The red part of the optical spectra may also be affected by the host
contamination, although not as much as in the NIR.  We also tried to
remove the host contribution when extracting the spectra as described 
in \S\ref{sec:uvopt}.

\subsection{Galactic Reddening Correction\label{sec:redden}}

The FUV-to--optical spectra suffer from Galactic dust extinction.
We corrected this with an empirical mean extinction law
\citep{Carde89}, assuming $R_v = A_V/E(B-V)=3.1$, a typical value for
the diffuse interstellar medium.  $E(B-V)$ is obtained from
NED\footnote{NASA/IPAC Extraglactic Database (NED) is operated by the
Jet Propulsion Laboratory, California Institute of Technology, under
contract with the National Aeronautics and Space Administration.} 
based on the dust map created by Schlegel et al. (1998).  

The FUV-UV-optical spectra are combined first before applying the
Galactic reddening correction.  The FUV spectra from \fuse\ extend
below 1000\,\AA, which the \citet{Carde89} extinction curve does not
cover.  \citet{Shang05} has shown a short extrapolation of the
extinction curve below 1000\,\AA\ is acceptable, and we use the same
technique here.


\subsection{Measurements}\label{sec:calculation}

We have made detailed measurements of all the spectral properties (continua
and emission lines), which need further analyses and will be presented
in a separate paper.  Here, we briefly describe the measurements and
derived quantities related to this work.  These quantities are listed
in Table~\ref{tb:sample}.

\subsubsection{Redshift\label{sec:redshift}}

Since all our objects have high-quality UV-optical spectra, we used
the optical narrow line \oiiil\ to define the rest frame of each object,
and double-checked against other strong narrow emission lines.  The centroid
of \oiiil\ is obtained by fitting this spectral region with a power-law
for the local continuum and Gaussian components for different emission lines
simultaneously \citep[see][for details]{Shang05,Shang07}.  
We can reach a redshift accuracy of 0.0002 for most objects.

When \oiiil\ is weak or missing from our spectral coverage for some
objects, we have obtained the redshift from NED as an initial guess in
our spectral fitting, checked against the fitted centroids of other
available strong emission lines, and made corrections when needed.
The redshift uncertainty in this case is about 0.001, sufficient for
SED work.

\subsubsection{Radio Loudness\label{sec:loudness}}

The traditional definition of radio loudness R is the ratio of
rest-frame flux density at radio 5\,GHz to that at optical 4400\,\AA,
$R = f(5GHz)/f(4400\mbox{\AA})$, and $R=10$ separates RL and RQ
objects.  We use $f(4215\mbox{\AA})$, instead of $f(4400\mbox{\AA})$,
because this local continuum is well defined (see
\S\ref{sec:msed}) in our spectra and it makes little difference in
calculating R.  To obtain $f(5GHz)$, we have interpolated for most
objects using two radio measurements embracing 5\,GHz (rest-frame) in
frequency.  For radio-quiet objects, there is usually only one
measurement around 5\,GHz in observed-frame, we therefore assume a
flat spectral index (in $f_\nu$) and take the value as rest-frame $f(5GHz)$ as
well.  Since all RQ objects have $z<0.5$, this will not cause a big
error, especially when using R to distinguish RL and RQ quasars.

\subsubsection{Luminosity\label{sec:lbol}}

The continuum luminosity is given as  $ \lllambda (3000\mbox{\AA})$, 
measured at 
3000\,\AA rest frame wavelength, and assuming a flat $\Lambda$CDM cosmology with $H_{0}=70$
km\,s$^{-1}$\,Mpc$^{-1}$, $\Omega_{m}=0.3$ and $\Omega_{\Lambda}=0.7$.
If desired, an average multiplicative correction factor of 5, taken
from \citet[][Fig.~12]{Richa06}, can be applied to  $ \lllambda
(3000\mbox{\AA})$, to estimate the bolometric luminosity.
Other more refined theoretical bolometric corrections can 
also be adopted from \citet[][]{NemBro10}.

We measure the bolometric luminosity for our individual quasars in
various ways and report the results in a forthcoming paper (Runnoe et
al. 2011).  On average, the bolometric luminosities are very similar
to 5$\lambda L_{\lambda3000}$.  There are a number of issues to
consider in making bolometric corrections and caution is advised.


\section{SEDs\label{sec:sed}}
\subsection{SEDs for Individual Objects}

\begin{figure}
\plotone{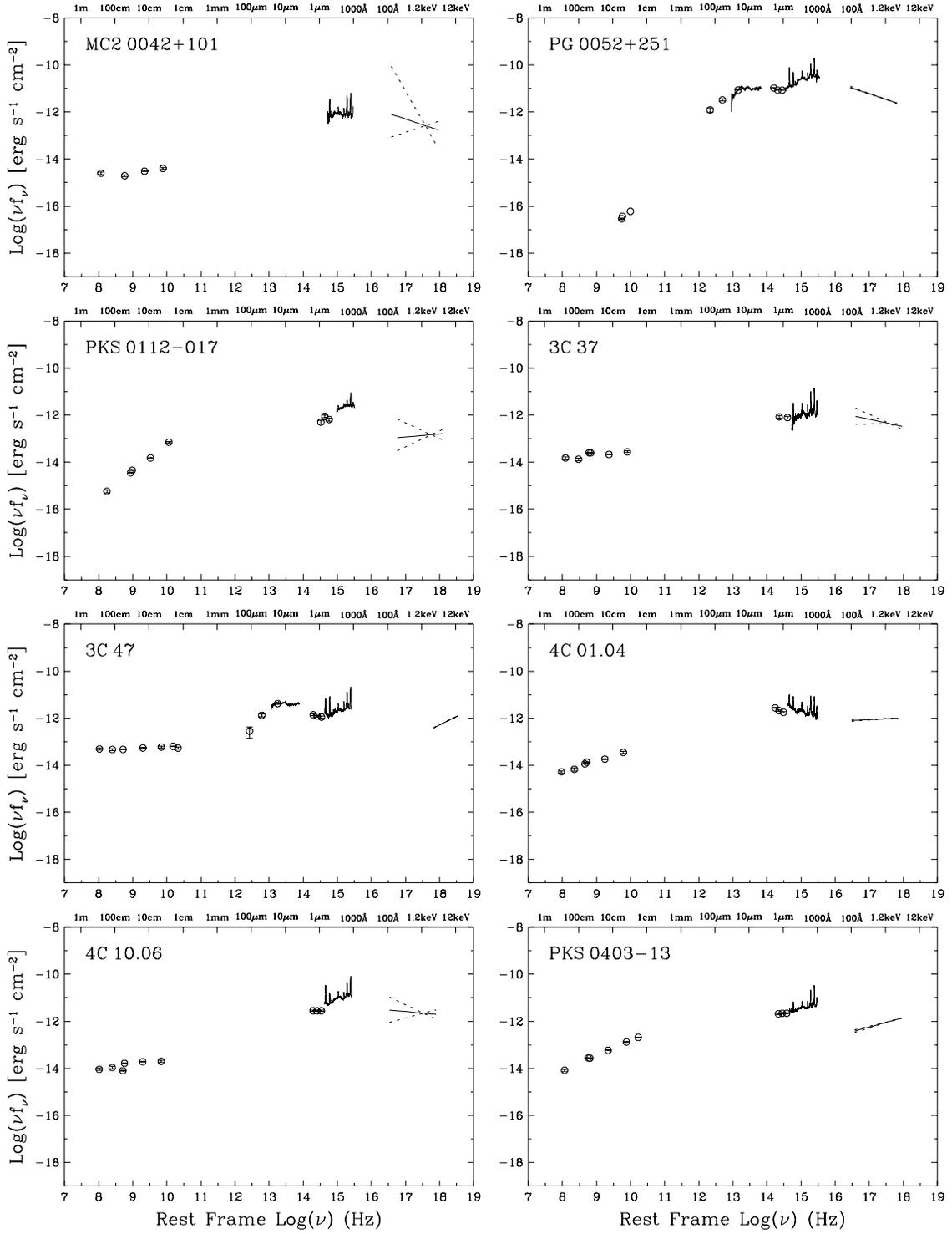}
\caption{SEDs of the quasar sample, ordered by RA to match
Table~\ref{tb:sample}.  The flux ($\nu f_\nu$) is in
observed-frame.
Figures~\ref{fg:allsed}.1-\ref{fg:allsed}.11 
are available in the online version of the Journal.
\label{fg:allsed}
}
\end{figure}

With multi-wavelength data in hand, it is very straightforward
to combine the data to build the SEDs for individual objects
(Fig.~{\ref{fg:allsed}}).  FUV-to-optical spectra are rebinned in
the observed-frame to a lower resolution, but not so much that the 
emission-line features are degraded too much.  
The bin size is 10\,\AA, corresponding to
1000\kms\, at 1000\,\AA, and 500\kms\, at 6000\,\AA.  Our \spitzer\ IRS
mid-IR spectra have low resolution and sampling of $\gtrsim
0.02\micron$, so we retain this sampling in the SEDs without invoking further
rebinning.  The re-built X-ray spectra
have a sampling of 0.1\kev (\S\ref{sec:xray}).

When we combined FUV-UV-optical spectra, we scaled data to photometric
nights or \hst\ observations \citep{Shang05, Shang07}.  When
photometric spectra overlapped, the agreement was better than 5\%
(e.g. between ground-based and \hst\ spectra, as well as
inter-compared optical spectra).

We present the data in \fnu\ vs.  frequency (Hertz) and
convert the flux density in each waveband to the same units of mJy
($10^{-26}$\,\fluxhz).  After combining all the data, we
apply a redshift correction to obtain the rest-frame SEDs.  Only
wavelength and frequency are shifted to the rest-frame and the flux
densities are left unchanged from the observed-frame.

As shown in Figure~{\ref{fg:allsed}}, while we try to collect
a uniform data set as much as possible, not all objects have data
in all wavelength bands except for the UV-to-optical.  
Table~\ref{tb:radio}--\ref{tb:xray} gives further information.

\subsection{Composite SEDs of RL and RQ Objects\label{sec:msed}}

One of the main motivations of this study is to update the mean quasar
SEDs of Elvis et al. (1994) using data from modern telescopes of
higher sensitivity and better resolution.  

We divided the sample into radio-loud and radio-quiet samples.  For
each sample, we first normalized the flux density of each object at
rest-frame 4215\,\AA\ where, after visual inspection of all spectra, 
there seems to be no strong emission features.
The actual normalization factor is
the mean flux density within 30\,\AA\ around 4215\,\AA.  The bandpass is
chosen to be small, to avoid emission features, and large enough to minimize
the noise in calculating the mean.
For seven higher redshift radio-loud objects, their rest-frame spectra
do not cover 4215\,\AA.  We therefore normalize them at 2200\,\AA, another
continuum region, to a composite spectrum built with all spectra
normalized earlier at 4215\,\AA\ in the same sample.  The
normalization factor is derived from the mean flux density within 50\,\AA\ around
2200\,\AA\ in this case.

After normalization, we visually check the distribution of all the
points from all objects and define the final bins in each waveband for
calculating the composite SEDs.  Each bin contributes one point in the
final composite SED and the central frequency of each bin
represents the final frequency of that point in the SEDs.

\begin{figure}
\includegraphics[angle=270,scale=0.6]{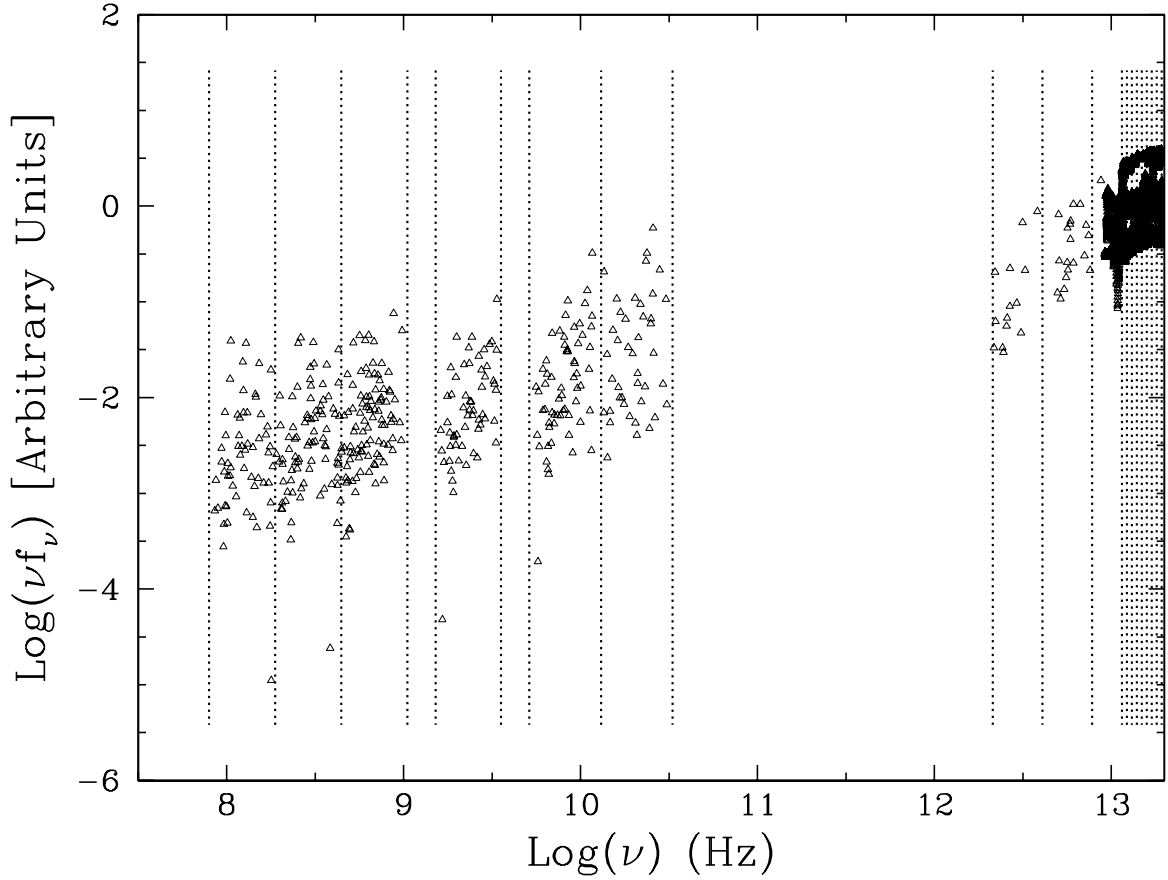}
\caption{Example of defining bins in constructing composite SEDs for the RL
sample.  Open triangles are all data points from all RL objects.
Vertical dotted-lines define the bins, each of which contributes one point
in the final composite SED.  Note the two small gaps with no data point
between Log($\nu$) of 9 and 10.  Shown on the right are the 2 FIR bins and
MIR spectra where the bin size is much smaller.
}
\label{fg:binsize}
\end{figure}

For each waveband with photometric points (radio, FIR, NIR), we locate
a logarithmic frequency range (rest-frame) to enclose all points and then
define a few bins with equal bin size within the range.  Since radio
data span a large frequency range, sometimes there are obvious gaps in
the distribution.  In such cases, we define more than one frequency
range to avoid the gaps and still try to keep similar bin size across
the ranges.  Figure~\ref{fg:binsize} shows an example of defining bins
for the RL sample.

For spectroscopic data (UV-optical, MIR, X-ray), it is easy to define
consecutive bins with the same bin size.  The bin size is chosen to have
enough points in each bin for statistical significance and still be
able to preserve the emission features.  Table~\ref{tb:bins} lists
the parameters we use to define the bins for each waveband.  

\begin{figure}
\includegraphics[angle=270,scale=0.6]{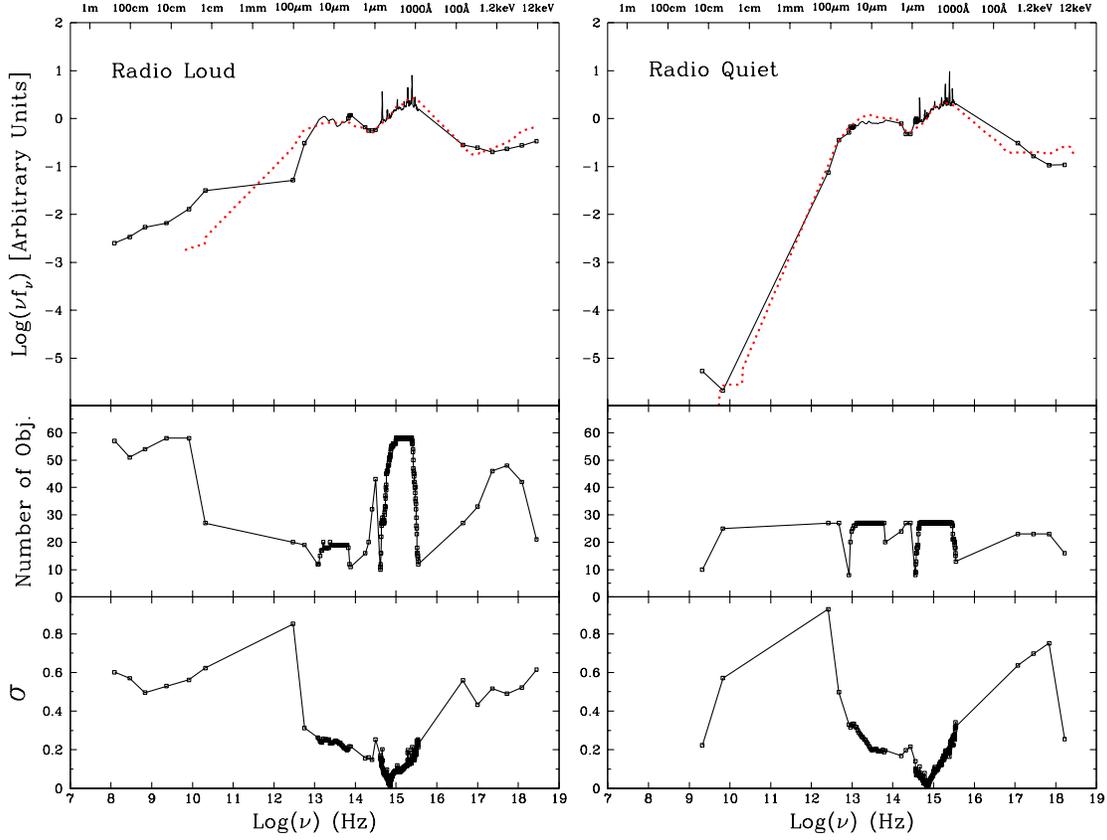}
\caption{{\bf Top:} Median SEDs for radio-loud and radio-quiet
samples. Overploted are the mean SEDs (dotted line) of Elvis et al.\
(1994), normalized at about 4000\,\AA.
{\bf Middle:} Number of objects used to construct the SEDs in each
wavelength bin marked with a small square.  There are no data outside the
bins.
{\bf Bottom:} Standard deviation around the mean in each wavelength bin.
The data of the median SEDs are available in the 
online version of the Journal.
}
\label{fg:msed}
\notetoeditor{Please display it for the full width of the page}
\end{figure}

After having defined the bins for a sample of interest, we rebin the
data to obtain one single value for each bin.  This is mostly
necessary for spectroscopic data, and this is done for each object
separately so that all objects with available data will have equal
weights in building the composite SED.
%
%
Two RQ objects have upper limits in the highest radio frequency bin
and six objects each for RL and RQ samples have upper limits in the
MIPS 160\micron\ band.  They are included in the median combining
process but none of the upper limits has a flux higher than the median
value in the corresponding bin, therefore their uncertainty does not
affect the composite SED.
This median combining is very effective in rejecting outliers and
preventing any extreme objects from dominating the final SEDs.  We
therefore also refer to our composite SEDs as median SEDs.  Finally,
if one bin has less than 8 points (i.e., objects), we exclude this bin
from the median SED.  Figure~\ref{fg:msed} shows our median SEDs for
RL and RQ samples.

%
%




\subsection{Discussion\label{sec:discussion}}

We try to keep all the original data in building the SEDs of individual
objects.  The only change is the re-sampling of the UV-optical spectra
to 10\,\AA\ resolution by rebinning.  Although we lose some useful
information (e.g. resolving narrow emission lines), 
this does not affect the SED work at all.  

In constructing the composite SEDs, we applied rebinning again mostly
for spectroscopic data.  We did not apply any smoothing or
interpolation in regions with real data, 
which could introduce systematic biases.  
The features in our median SEDs are real.


At the edges of some wavebands, the number of objects with data drops
sharply (Fig.~\ref{fg:msed}, middle), and our method of using median to build
the composite SEDs can help to some extent in the small number statistics.
We also visually check to ensure that the SEDs are reasonably smooth
in those regions.

\subsubsection{RL vs. RQ\label{sec:lowzrl}}

\begin{figure}
\includegraphics[angle=270,scale=0.6]{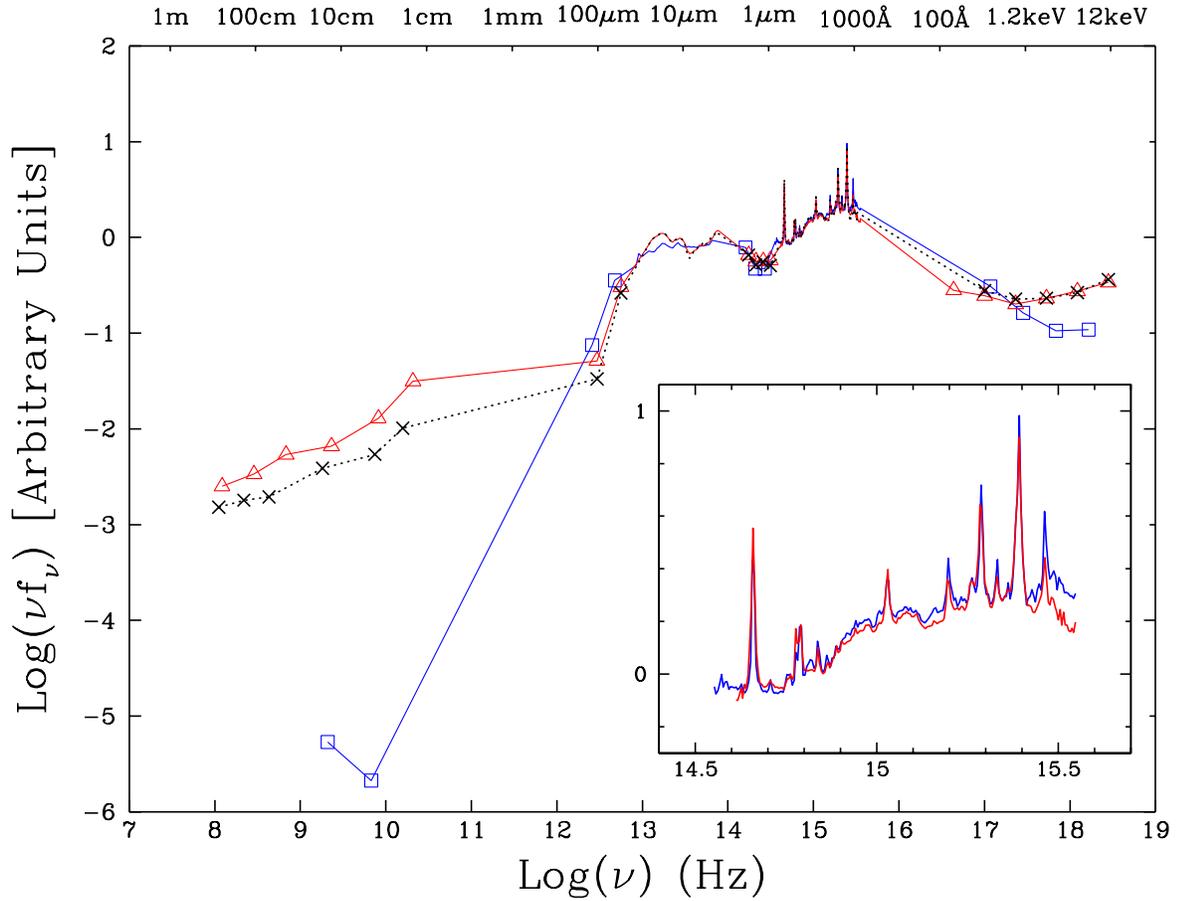}
\caption{Median SEDs for the entire radio-loud sample (open
triangle, red) and and radio-quiet sample (open square, blue).
Also shown is the median SED for only 
the low-redshift radio-loud objects (cross, dotted-line) for
comparison (\S\ref{sec:lowzrl}).  The inset shows only the UV-optical
region.
}
\label{fg:lowzrl}
\notetoeditor{Please display it for the full width of the page}
\end{figure}

Figure~\ref{fg:lowzrl} overplots our median SEDs of RL and RQ samples,
normalized at 4215\,\AA\ during the construction.  The SEDs from FIR
to UV are very similar for RL and RQ, especially in the UV-optical
region.  This is only true for the UV-optical continuum, because
emission lines, such as \feii\ and \oiii, are known to strongly
correlate with radio loudness in the Eigenvector I relationship
\citep{BG92}.  We will investigate the relationships between SEDs and
emission line properties in a future paper.

The biggest difference between RL and RQ median SEDs is in
the radio, where luminosity could differ by 3 orders of magnitude.
There is also an obvious difference in the X-ray, where the RL objects
are more X-ray
luminous.  This correlation between radio and X-ray luminosity has
been reported in previous studies \citep[e.g.,][]{Brink00,Polle07}.

During the construction of the median SEDs, we have also defined
6 bins in the radio for the RQ sample 
(Table~\ref{tb:bins}), however there are not
enough objects ($\geq 8$) in 4 of the bins.  Therefore, there are only
two points good enough to be included in the final radio SED of RQ
sample.  Given the big difference in the number of objects involved in
these two bins (Fig.~\ref{fg:msed}), the apparent steep RQ spectral
index, if defined using the two points, may not be reliable.

Our RL sample has more objects with higher redshifts than the RQ sample
(Fig.~\ref{fg:hist}).  We therefore build another median SED of RL
objects only with redshift less than 0.5 (21 objects), similar to
those in the RQ sample.  Comparing this and the SED of the entire RL
sample (Fig.~\ref{fg:lowzrl}), we find that the only notable
difference is in the radio where the low-z subsample shows a less
luminous radio SED, but radio spectral index seems similar.
The difference is probably real because of sample properties ---
high-z RL objects are more luminous in radio, but the differences
between the RL and RQ samples are still much more prominent.


\subsubsection{Comparison with E94 Mean SEDs}

\citet{Elvis94} use 47 objects to build the mean SEDs (MSED94) for RL
and RQ objects.  There are 11 objects in common with our sample,
including 6 RL and 5 RQ objects.  We compare our median SEDs of RL and
RQ quasars (Fig.~\ref{fg:msed}) with MSED94.  The overall shape of the
SEDs over the entire available frequency range is similar, but
there are more detailed features in our new SEDs.

We specifically keep all the emission features in the UV-optical
region because they are real spectral features and our data quality
allows us to keep them.  The underlying continuum shapes in this
region look similar to those of MSED94.  We note that our UV-optical
SEDs extend to shorter wavelength beyond 1000\,\AA, and start to turn
over, indicating the peak of the ``big blue bump''
\citep[e.g.,][]{Zheng98,Shang05}.  This is especially obvious in our RL median SED
where we have more higher redshift objects.

In the MIR, the broad silicate emission features around 10 and 18\micron\
are prominent in the SEDs.  These could not otherwise be reproduced
without the unprecedented spectral data from \spitzer\ IRS.  To the
shorter wavelength of these features, a well-defined power-law rises
up to about 4\micron, the IRS detecting limit for our sample in the
rest-frame.  However, the well-known inflection around 1\micron\ is also
well defined by the red optical spectroscopy and NIR 2MASS photometry.  It
is therefore very clear that somewhere between 1 and 4\micron, there
is an infrared bump, which is further supported by the fact that the NIR
K-band data point starts to rise toward MIR in both RL and RQ SEDs.
Other studies have suggested that there is a 3\micron\ bump, resulting from
the hottest dust in AGNs \citep{Netze07,Hiner09}.


Although MSED94 have a lot of upper limits in the FIR from IRAS while
we use the latest \spitzer\ MIPS data, the SEDs agree surprisingly well
for RQ sample in the FIR and extending to the radio.

For the RL sample, it is expected that our \spitzer\ data define a better FIR SED,
which falls more steeply toward longer wavelengths.  Our radio SED is more
luminous than that of MSED94, simply because there are more
radio-luminous objects in our sample.

\subsubsection{Comparison with Quasar SEDs of \citet{Richa06} }

\begin{figure}
\epsscale{0.8}
\includegraphics[angle=270,scale=0.6]{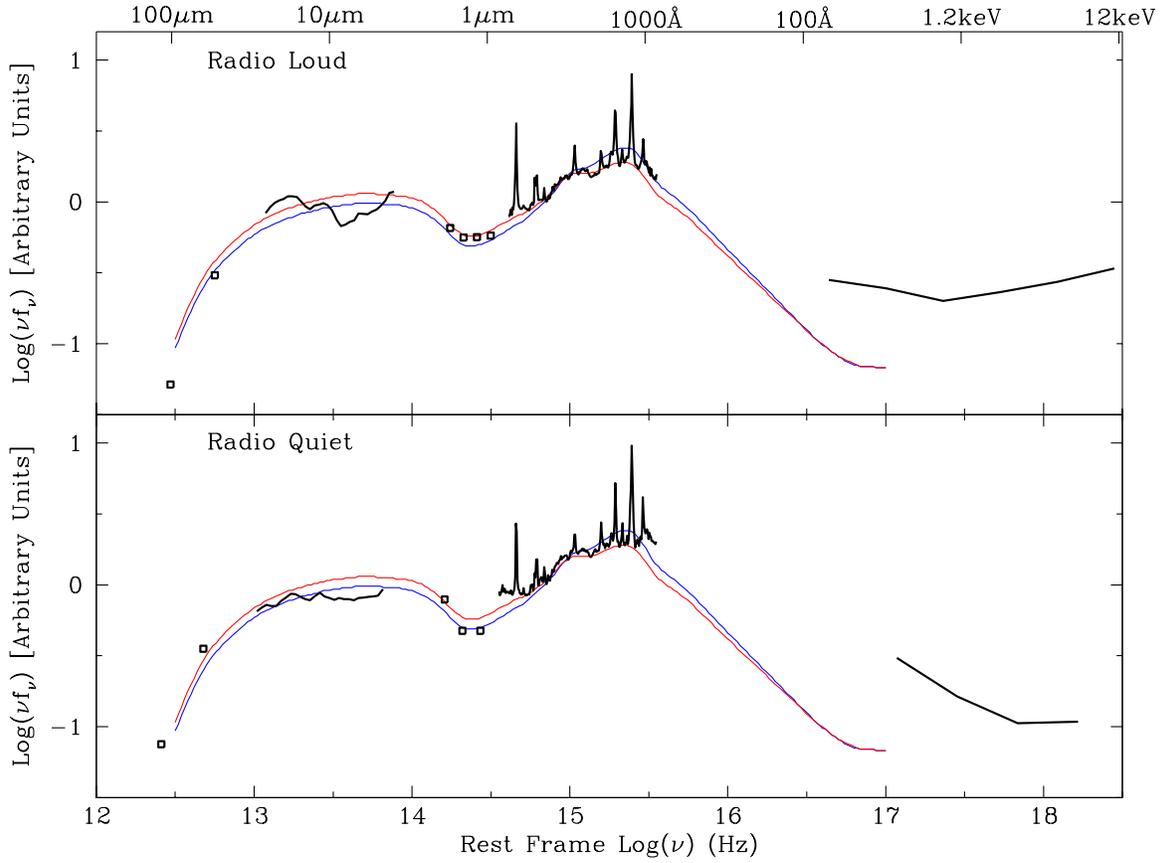}
\caption{Median SEDs for our sample (black) compared
with the SEDs of \citet{Richa06} which are essentially for the radio-quiet
quasars since there are only 8 radio-loud objects involved.
The red and blue lines are for the optically red and blue halves of
the population, respectively.
The SEDs are normalized at 4200\,\AA.
}
\label{fg:compareR06}
\notetoeditor{Please display it for the full width of the page}
\end{figure}

We compare our SEDs with those of \citet[][hereafter R06]{Richa06} in
Figure~\ref{fg:compareR06}.  R06 has provided the broadest frequency
coverage, from Far-IR to X-ray, in recent SED studies, and their
sample of 259 SDSS quasars extends to higher redshift and higher
luminosity than ours.  They constructed the SEDs using photometry
points, including 5 SDSS magnitudes and 4 \spitzer\ InfraRed Array
Camera (IRAC) flux densities, supplemented by available \galex\ $f$
and $n$ bands, $J$, $H$, and $K$, the \iso\ 15\,\micron\ band, and the
\spitzer\ MIPS 24 and 70\,\micron\ bands.  When objects do not have
measurements in the supplemental bands, they used \citet{Elvis94} SEDs
and their ``gap repair'' technique to estimate the missing values.
Their X-ray fluxes were obtained from \rosat\ detections.  When no
detection is available, they estimated the X-ray flux using the
2500\,\AA\ flux and the tight $L_{UV}(2500\mbox{\AA})$-$L_{X}$
relationship \citep{Strat05}.  They have only 8 radio-loud quasars, so
the final SEDs are essentially for radio-quiet quasars.  Therefore, we
only compare our radio-quiet SED with theirs, but plot both of our RL
and RQ SEDs in Figure~\ref{fg:compareR06} for completeness.  We
further choose to compare only with their two SEDs constructed with
optically red and blue halves of their sample, and their mean SED
would lie between the two.

Our SEDs show relatively much higher X-ray emission, indicating that
our sample (and that of \citet{Elvis94}) is not representative of the
SDSS quasars in this region and probably does not follow the 
$L_{UV}(2500\mbox{\AA})$-$L_{X}$ relationship found in SDSS quasars
\citep{Strat05}.
In the Far-IR to near-IR region, the overall shape and trend seem to
match very well, but ours have more detailed features.

Although the SEDs are normalized at 4200\,\AA\
(Fig.~\ref{fg:compareR06}), their shapes also match well over most of
the UV-optical region.  Only at the two ends, our SED show optical
redder and UV brighter.  This implies the different sample properties,
because our objects are mostly UV-bright quasars and are probably
redder in the $i$ band compared with R06 sample.  Many objects in the
R06 sample have much higher redshifts and luminosity than ours, but we
do not have enough information to address any possible evolution or
luminosity-dependent effects in quasar SEDs by comparing them.

\subsubsection{Comparison with Other Quasar Composites}

\begin{figure}
\epsscale{0.8}
\includegraphics[angle=270,scale=0.6]{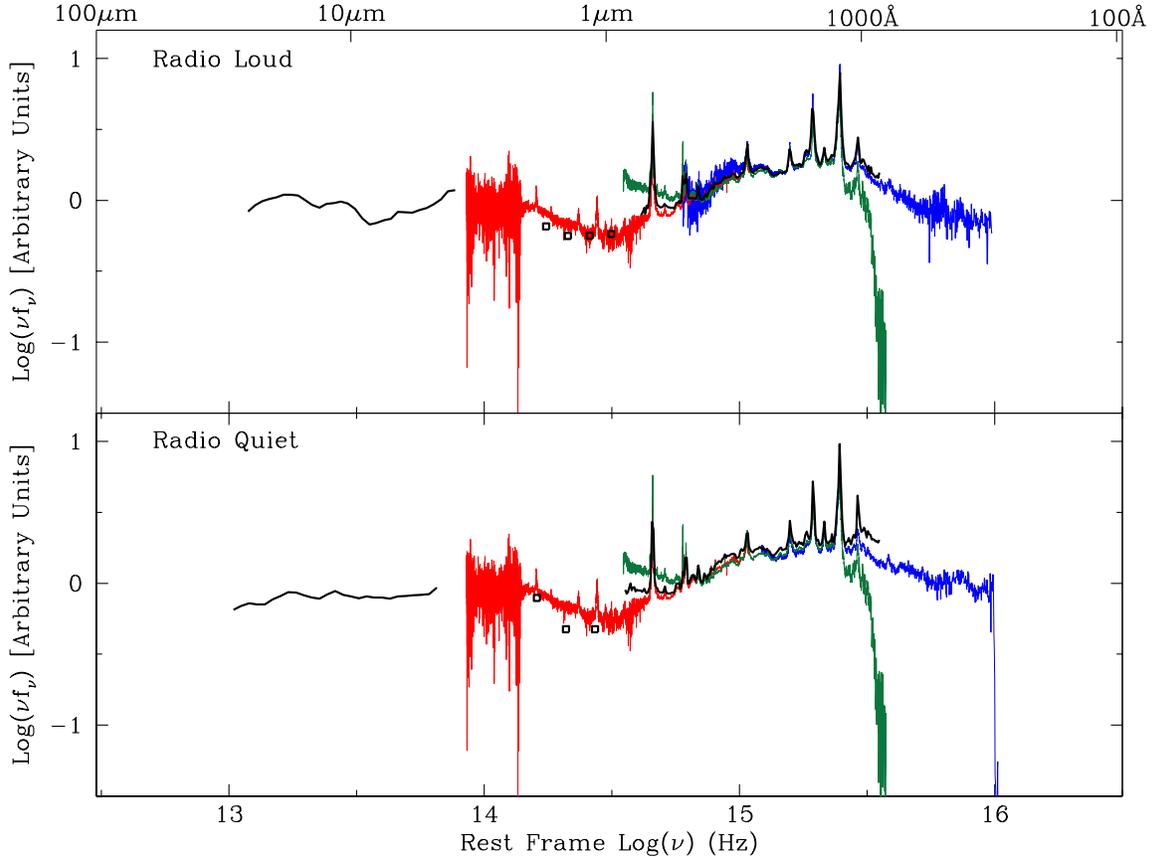}
\caption{Median SEDs from FUV to MIR for our sample (black) compared
with other quasar composite spectra, including the HST ultraviolet
composites \citep[blue,][]{Telfe02}, the SDSS composite
\citep[green,][]{Vande01}, and an NIR composite of
27 low-redshift SDSS quasars \citep[red,][]{Glikm06}.  The SDSS and NIR
composites do not distinguish radio-loud or radio-quiet objects.  All
composites are normalized at 4200\,\AA.
}
\label{fg:compare}
\notetoeditor{Please display it for the full width of the page}
\end{figure}

Figure~\ref{fg:compare} shows the comparison between our SEDs from
Far-UV to Mid-IR and other quasar composite spectra, including the HST
ultraviolet composites \citep[][]{Telfe02}, the composite from Sloan
Digital Sky Survey \citep[SDSS,][]{Vande01}, and a Near-IR composite
of 27 SDSS quasars \citep{Glikm06}.

The HST composites extend beyond the FUV part of our SEDs to the
extreme-UV (EUV), revealing the EUV peak more clearly.  In the
overlapping region, our SEDs are in good agreement with the HST
composites although the radio-quiet HST composite drops a little
steeper at higher frequency.  The SDSS composite includes both
radio-loud and radio quiet objects and is consistent with our SEDs
between about 1200 and 4500\,\AA.  The increased flux at the red part
of the SDSS composite is partly due to host galaxy contamination at
low redshift, while the blue part beyond \lya\ should be ignored
because of \lya\ forest contamination.  The NIR composites of
radio-loud and radio-quiet samples show little difference in
\citet{Glikm06}.  We use the composite of their entire sample and its
optical part matches our SEDs well.  Its NIR region shows the
1\micron\ inflection as expected, and the overall shape connects our
MIR and optical SEDs surprisingly well.  Although our NIR points are
located a little lower than the composite, the continuum trends are
consistent.

\subsection{Uncertainty and Caveats}

As in all previous quasar SED studies \citep[e.g.,E94; R06][]{},
the dispersion of the mean or composite SEDs is large.  The
dispersion of our median SEDs can be evaluated in Fig.~\ref{fg:msed},
where population standard deviations of all rebinned points
in each bin are plotted at the bottom.
Because we normalized the individual SEDs at 4215\,\AA,
the dispersion is minimized in the optical and increases toward both
low and high frequencies to about 0.6 (dex) in radio and X-ray.  Even
in the NIR to UV region, the standard deviation increases rapidly away
from the normalization wavelength.  

\begin{figure}
\epsscale{0.8}
\includegraphics[angle=270,scale=0.6]{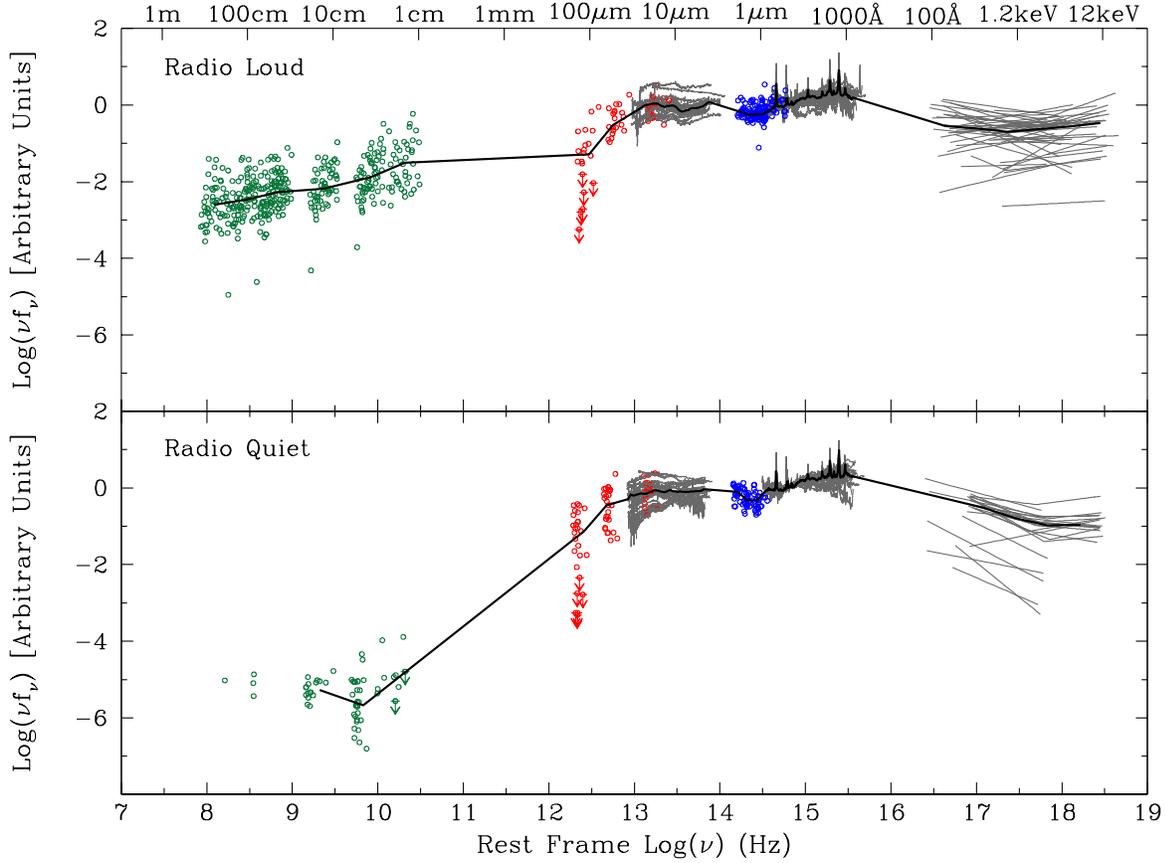}
\caption{Dispersion of the median SEDs.
All SEDs of individual objects, normalized at 4215\,\AA, are plotted
together, showing large dispersions about the median SEDs (black thick
line).  Open circles are for radio (green), FIR (red), and NIR
(blue) data. Spectroscopic data are shown in grey for MIR, UV/optical, and
X-ray.  
}
\label{fg:dispersion}
\notetoeditor{Please display it for the full width of the page}
\end{figure}

We also show the dispersion in Figure~\ref{fg:dispersion}, where we plot
all data from normalized SEDs of individual objects and the median
SEDs built from them.  The actual difference between the individual SEDs can
be more than 2 orders of magnitude.  Even in the NIR to UV, the
difference is still more than one order of magnitude.  These all
indicate the large variation of quasar SEDs.

We note a caveat that the NIR host galaxy corrections may have a large
uncertainty (\S\ref{sec:host}, Table~\ref{tb:hostfrac}), because the
H-band host galaxy measurements with ground-based infrared data may be
somewhat inherently uncertain.  The host galaxy fractions from HST
observations \citep[][and our own observations]{McLMcL01}are
systematically smaller than those from ground-based observations
\citep[][and our IRTF data]{McLRie94a,McLRie94b}.  Two objects have
both ground-based and HST H-band observations and show revised lower
host galaxy fractions from HST data (Table~\ref{tb:hostfrac}), with
PG~1322+659 from 43\% to 22\%, and PG~1427+480 from 45\% to 26\%,
respectively.
In addition, although the same technique was used in estimating the
host galaxy fraction with all ground-based data, host galaxy fractions
from \citet[][]{McLRie94a} and our IRTF data seem to be systematically
higher than \citet[][]{McLRie94b} by about 20\%.  Some SEDs show
discrepancy between NIR broad-band points and optical spectra
(Fig.~\ref{fg:allsed}), indicating possible over-subtraction of host
galaxy contributions.  Readers should be aware of this issue when
using data of this region for their own applications.

Our sample is heterogeneous, and there is the possibility that the
sample may be biased.  \citet{Jeste05} pointed out that BQS quasars
are representative of bright blue quasars, but not representative of
bright red quasars. Since our sample involves many PG quasars, it is
possible that we might be missing some red quasars, and therefore the
SEDs are not truly representative of $all$ UV-bright quasars. If this
is true, it will affect the spectral index in the optical region.
This does not seem to be a serious problem, however, because our SEDs
do not show a drop off in the red part of the optical region when
compared with the SEDs of R06 (Fig.~\ref{fg:compareR06}).  
Moreover, even for the distinguished blue and red quasars in R06,
except for the different spectral indices in the optical, their mean
SEDs are very similar (see Fig.~\ref{fg:compareR06}, or their
Fig.~11), the largest difference in $log(\nu f_\nu)$ is about 0.1 dex,
significantly smaller than the dispersion of either R06 or our
composite SEDs.  Therefore, the large variation of individual quasar
SEDs is still the dominant factor for the uncertainty of the composite
SEDs.


\section{SUMMARY AND FUTURE WORK\label{sec:summary}}

We have compiled SEDs for 85 quasars using high-quality
multi-wavelength data from radio to X-ray energies.  The data were obtained
from the next generation space telescopes and ground-based telescopes.
Using these data, we have constructed composite (median) SEDs for
radio loud and radio quiet quasars.  This work is an update on the
mean SEDs built by \citet{Elvis94} with about twice as many objects.
On our website\footnote{\url{http://physics.uwyo.edu/agn/}} and the
online version of the Journal,
we make the electronic version
of the median SEDs available for public use.  We
caution again that, because of the large variation in quasar SEDs, any
composite SEDs should be used with care.  Our composite SEDs are 
representative only for UV-optical bright quasars.  The RQ median SED is
constructed from low-redshift ($z<0.5$) objects, while the RL median SED comes
from objects of redshift up to 1.4.

We also plan to investigate
the multi-wavelength data of individual objects.
We have measured all the spectral parameters of
the entire sample and the work will be presented in a separate paper.
We will be able to obtain the bolometric luminosities from real data and
investigate the bolometric corrections associated with spectral properties, such
as continua or emission features.  
We have also planned to investigate how the quasar SED varies with
different physical parameters such as black hole mass and Eddington
ratio.
The
multi-wavelength data will help us to better understand the quasars. 
A series of papers based on this data set are forthcoming.

\acknowledgments

This work has been supported by the National Natural Science
Foundation of China through Grant No. 10773006 and Chinese 973 Program
2007CB815405.  We are also grateful for support by NASA through grant
HST-GO-10717.01-A, Spitzer-GO-20084, and Grant No. NNG05GD03G.  B. J.
W.  acknowledges financial support by NASA through LTSA grant
NAG5-3431 and HST grant GO-2578.01.87A, GO-4504, GO-5441, GO-06781,
and AR-5820 from the Space Telescope Science Institute, which is
operated by the Association of Universities for Research in Astronomy,
Inc., under NASA contract NAS5-26555.  We are also grateful for
support from the US National Science Foundation, through Grant No.
AST-0206261, AST-8794137 (B. J. W.) and AST-0507781 (MSB).

This publication makes use of data products from \vla, \spitzer,
\tmass, \hst, \fuse, \chandra, \xmm, \rosat, and other facilities.
We thank all those who built and operate them, who conducted
the surveys,  and who made available the data products and catalogs
cited in this paper.

\clearpage

%
%
%
%
%
%
%
%
%

\clearpage

\clearpage

\clearpage

\clearpage

\begin{deluxetable}{rlllllcrcc}
\tabletypesize{\tiny}
\tablecolumns{10}
\tablewidth{0pc}
\tablecaption{SED Sample \label{tb:sample}}
\tablehead{
\colhead{ID} &
\colhead{Name} & \colhead{Other Name}  & \colhead{RA(J2000)}& 
\colhead{DEC(J2000)} &
\colhead{$z$\tablenotemark{a}}
& \colhead{E(B$-$V)\tablenotemark{b}}
& \colhead{R\tablenotemark{c}}
& \colhead{$\lllambda (3000\mbox{\AA})$\tablenotemark{d}}
& \colhead{SampleID} 
}
\startdata
 1& MC2 0042+101&  
	&00:44:58.72 & $+$10:26:53.7 & 0.5870 & 0.068
	&   840\phm{.13} & 46.22 & RLQ\\
 2& PG 0052+251&  
	&00:54:52.10 & $+$25:25:38.0 & 0.1544 & 0.047
	&  0.34 & 46.30 & FUSE\\
 3& PKS 0112$-$017& UM 310
	&01:15:17.10 & $-$01:27:04.6 & 1.3743 & 0.062
	&  2819\phm{.13} & 45.92 & RLQ\\
 4& 3C 37&  
	&01:18:18.49 & $+$02:58:06.0 & 0.6661 & 0.039
	&  5550\phm{.13} & 45.64 & RLQ\\
 5& 3C 47&  
	&01:36:24.40 & $+$20:57:27.0 & 0.4250 & 0.061
	&  6570\phm{.13} & 44.94 & RLQ\\
 6& 4C 01.04& PHL 1093
	&01:39:57.25 & $+$01:31:46.2 & 0.2634 & 0.029
	&  2556\phm{.13} & 45.93 & RLQ\\
 7& 4C 10.06& PKS 0214+10
	&02:17:07.66 & $+$11:04:10.1 & 0.4075 & 0.109
	&   472\phm{.13} & 45.54 & RLQ\\
 8& PKS 0403$-$13&  
	&04:05:34.00 & $-$13:08:13.7 & 0.5700 & 0.058
	&  5413\phm{.13} & 46.21 & RLQ\\
\enddata
\tablenotetext{a}{Redshift, measured using our data (\S\ref{sec:redshift}).}
\tablenotetext{b}{From NED \url{(http://nedwww.ipac.caltech.edu/)} based on 
\citet{Schl98}.} 
\tablenotetext{c}{Radio loudness, $R = f(5\,GHz)/f(4215\mbox{\AA})$, 
calculated using our data (\S\ref{sec:loudness}).}
\tablenotetext{d}{Rest-frame luminosity at 3000\,\AA, 
calculated using our data (\S\ref{sec:lbol}).}
\end{deluxetable}

\begin{deluxetable}{rlcrc}
\tabletypesize{\scriptsize}
\tablecolumns{5}
\tablewidth{0pc}
\tablecaption{Radio Surveys
\label{tb:radiosurvey}
}
\tablehead{
\colhead{Frequency} & & \colhead{Number of} 
&\colhead{Resolution}
& \\
\colhead{(MHz)}&\colhead{Surveys} &\colhead{Objects Included} 
&\colhead{(\arcsec)}
&\colhead{References}
}
\startdata
74      & VLSS		  & 57    & 45		& 1 \\
151     & 7C		  & 30    & 70		& 2 \\
178     & 4C		  & 44    & 1380-1860	& 3 \\
325     & WENSS		  & 29    & 54		& 4 \\
365     & TEXAS		  & 55    & 22.1	& 5 \\
408     & MRC		  & 23    & 157		& 6 \\
1400    & NVSS/FIRST	  & 64/3  & 45/5	& 7 \\
1420    & Ulvestad.2005   & 2     & $<0.01$	& 8 \\
\enddata
\tablecomments{Radio surveys and references used to collect SED data.
}
\tablerefs{
(1) The VLA Low-Frequency Sky Survey \citep[][]{Cohen07}; 
(2) 7C 151-MHz Survey \citep[][]{Hales07};
(3) 4C Survey \citep[][]{Pilki65,Gower67};
(4) The Westerbork Northern Sky Survey \citep[][]{Renge97,Leide98};
(5) The Texas Survey of Radio Sources \citep[][]{Dougl96};
(6) The Molonglo Reference Catalogue of Radio Sources 
	\citep[][]{Large81,Large91};
(7) The NRAO VLA Sky Survey \citep[][]{Condo98}; 
   Faint Images of the Radio Sky at Twenty-Centimeters
    \citep[][]{Becke95};
(8) \citet[][]{Ulves05};
(9) \citet[][]{Barva96};
(10) \citet[][]{Leips06};    
(11) GB6	      
(12) The Parkes-MIT-NRAO \citep[][]{Griff95};
(13) \citet[][]{Kelle89};
(14) \citet[][]{Gear94};       
(15) \citet[][]{Barva05};
(16) \citet[][]{Kelle73};
(17) \citet[][]{Genze76};
(18) \citet[][]{Bolto04}.
}
\end{deluxetable}



\begin{deluxetable}{rlllll}
\tabletypesize{\scriptsize}
\tablecolumns{6}
\tablewidth{0pc}
\tablecaption{Radio Fluxes
\label{tb:radio}
}
\tablehead{
\colhead{ID}
&\colhead{Object}
&\colhead{$\nu$(MHz)}
&\colhead{$f_{\nu}$(mJy)}
&\colhead{$\Delta f_{\nu}$ (mJy)}
&\colhead{Reference}
}
\startdata
1& MC2 0042+101 &	      74 & 3380 & 390 & VLSS\\
			&&   365 & 540 & 47 & TEXAS\\
			&&  1400 & 218.8 & 7.0 & NVSS\\
			&&  4850 & 83 & 8.1 & GB6\\
2& PG 0052+251 &	    4800 & 0.61 & 0.03 & Leipski.2006\\
			&&  5000 & 0.74 &  & Kellermann.1989\\
			&&  8600 & 0.7 &  & Barvainis.2005\\
3& PKS 0112$-$017 &	      74 & 780 & 120 & VLSS\\
			&&   365 & 974 & 25 & TEXAS\\
			&&   408 & 1110 & 70 & MRC\\
			&&  1400 & 1076.2 & 32.3 & NVSS\\
			&&  4850 & 1437 & 75 & PMN\\
\enddata
\tablecomments{
Table~\ref{tb:radio} is published in its entirety in the 
electronic edition of \apjs.
A portion is shown here for guidance regarding its form and content.
}
\end{deluxetable}




\begin{deluxetable}{rlrrr}
\tabletypesize{\tiny}
\tablecolumns{5}
\tablewidth{0pc}
\tablecaption{\spitzer\ MIPS FIR Fluxes
\label{tb:mips}
\label{tb:fir}
}
\tablehead{
\colhead{}& \colhead{} & \multicolumn{3}{c}{Flux (mJy)}
\\ \cline{3-5}
\colhead{ID}
&\colhead{Object}
&\colhead{24\micron}
&\colhead{70\micron}
&\colhead{160\micron}
}
\startdata
2& PG 0052+251 &
  $    70. \pm     3.4$ &  $   76.1 \pm     9.1$ &  $   65.1 \pm     14.$ \\
5& 3C 47 &
  $   33.5 \pm    2.35$ &  $    31. \pm    5.14$ &  $   15.2 \pm    7.58$ \\
10& 3C 175 &
  $   12.3 \pm    1.43$ &  $   25.3 \pm    5.06$ &  $<    2.5$ \\
11& 3C 186 &
  $   6.79 \pm    1.06$ &\nodata 	&\nodata 	\\
13& IRAS F07546+3928 &
  $   220. \pm    6.03$ &  $   154. \pm    11.5$ &  $<   1.08$ \\
14& 3C 207 &
  $   11.9 \pm     1.4$ &  $   23.7 \pm    4.56$ &  $   57.8 \pm    13.1$ \\
15& PG 0844+349 &
  $   92.4 \pm    3.91$ &  $   63.2 \pm    7.93$ &  $   107. \pm    18.1$ \\
20& PG 0947+396 &
  $   49.8 \pm    2.87$ &  $   117. \pm    10.2$ &  $   111. \pm    18.3$ \\
\enddata
\tablecomments{The upper limits for 160\,\micron\ are 3$\sigma$
limits, where $\sigma$ is the standard deviation of the local sky background.
}
\end{deluxetable}



\begin{deluxetable}{rllllclllc}
\tabletypesize{\tiny}
\tablecolumns{10}
\tablewidth{0pc}
\tablecaption{Near-IR JHK Magnitudes from 2MASS
\label{tb:2massmag}
}
\tablehead{
\colhead{}& \colhead{} & \multicolumn{3}{c}{Profile Fitting Magnitude}
&\colhead{}  & \multicolumn{3}{c}{Aperture Magnitude} & \colhead{}
\\ \cline{3-5} \cline{7-9}
\colhead{ID}&\colhead{Object}
&\colhead{J}&\colhead{H}&\colhead{K}
&&\colhead{J}&\colhead{H}&\colhead{K}
&\colhead{Ext\tablenotemark{a}}
}
\startdata
2& PG 0052+251 	& 13.891$\pm$0.032& 13.174$\pm$0.032& 12.239$\pm$0.022
		&& 13.813$\pm$0.022& 13.071$\pm$0.018& 12.194$\pm$0.023&    \\
3& PKS 0112$-$017 	& 16.935$\pm$0.172& 15.810$\pm$0.131& 15.664$\pm$0.212
		&& 16.311$\pm$0.183& 15.654$\pm$0.269& 15.087$\pm$0.321&    \\
4& 3C 37 	& 16.734$\pm$0.154& 15.913$$     & 15.099$\pm$0.136
		&& 16.528$\pm$0.400&\nodata$$     & 15.248$\pm$0.281&    \\
5& 3C 47 	& 16.352$\pm$0.112& 15.445$\pm$0.111& 14.553$\pm$0.093
		&& 16.086$\pm$0.212& 15.248$\pm$0.219& 14.339$\pm$0.093&    \\
6& 4C 01.04 	& 15.835$\pm$0.083& 14.866$\pm$0.082& 13.822$\pm$0.058
		&& 15.204$\pm$0.154& 14.459$\pm$0.105& 13.373$\pm$0.138& ext\\
7& 4C 10.06 	& 15.351$\pm$0.051& 14.547$\pm$0.049& 13.796$\pm$0.050
		&& 15.279$\pm$0.061& 14.572$\pm$0.103& 13.775$\pm$0.086&    \\
8& PKS 0403$-$13 	& 15.613$\pm$0.062& 14.822$\pm$0.081& 14.122$\pm$0.065
		&& 15.534$\pm$0.118& 14.186$\pm$0.155& 13.938$\pm$0.226& ext\\
9& 3C 110 	& 14.875$\pm$0.036& 14.733$\pm$0.050& 13.937$\pm$0.048
		&& 14.875$\pm$0.059& 14.853$\pm$0.133& 13.902$\pm$0.170&    \\
\enddata
\tablecomments{The 2MASS magnitudes are obtained from both profile
fitting photometry and aperture photometry.
Values without uncertainties reflect the fact that the original
sources did not provide uncertainties.
}
\tablenotetext{a}{An ``ext'' indicates that the aperture magnitude 
is from the 2MASS extended source catalog, instead of point source
catalog.
}
\end{deluxetable}

\begin{deluxetable}{rllllcllll}
\tabletypesize{\scriptsize}
\tablecolumns{10}
\tablewidth{0pc}
\tablecaption{Total JHK Magnitudes and Observed Host Fraction
\label{tb:2masshost}
\label{tb:hostfrac}
}
\tablehead{
\colhead{}& \colhead{} & \multicolumn{3}{c}{Total Mag}
&\colhead{}  & \multicolumn{3}{c}{Host Fraction} & \colhead{}
\\ \cline{3-5} \cline{7-9}
\colhead{ID}&\colhead{Object}
&\colhead{J}&\colhead{H}&\colhead{K}
&\colhead{}
&\colhead{J}&\colhead{H}&\colhead{K}
&\colhead{Reference\tablenotemark{a}}
\\
\colhead{(1)} & \colhead{(2)} & \colhead{(3)} & \colhead{(4)}
& \colhead{(5)} & \colhead{} & \colhead{(6)} & \colhead{(7)} & 
\colhead{(8)} & \colhead{(9)} 
}
\startdata
2& PG 0052+251 	& 13.81& 13.07& 12.19&& 0.27&0.22& 0.14& P2\\
13& IRAS F07546+3928 	& 12.89& 11.97& 11.00&& 0.73&0.53& 0.29& IRTF\\
15& PG 0844+349 	& 13.33& 12.67& 11.96&& 0.58&0.542& 0.36& P1\\
20& PG 0947+396 	& 14.42& 13.72& 12.64&& 0.33&0.27& 0.16& P3\\
21& PG 0953+414 	& 14.18& 13.46& 12.51&& 0.29&0.23& 0.15& P2\\
24& PG 1001+054 	& 14.99& 14.18& 13.03&& 0.75&0.58& 0.30&
IRTF (P1, $<0.273$)\\
28& PG 1100+772 	& 14.45& 13.75& 12.96&& 0.27&0.223& 0.18& HST\\
31& PG 1114+445 	& 14.15& 13.43& 12.29&& 0.54&0.461& 0.23& P1\\
\enddata
\tablecomments{Total magnitude is adopted from 2MASS aperture
magnitude in Table~\ref{tb:2massmag}.
}
\tablenotetext{a}{
Reference for adopted H-band host fraction.  
A reference ID and a number in 
a parenthesis indicate
another H-band host fraction from the corresponding reference.
P1---McLeod \& Rieke 1994a; P2---McLeod \& Rieke 1994b; P3---McLeod \&
McLeod 2001; IRTF---Our own observations from IRTF; HST---Our own
observations from HST NICMOS.  
}
\end{deluxetable}

\begin{deluxetable}{rllllcrrrl}
\tabletypesize{\scriptsize}
\tablecolumns{10}
\tablewidth{0pc}
\tablecaption{Host Corrected JHK Magnitude and Flux 
\label{tb:2masscorr}
}
\tablehead{
\colhead{}& \colhead{} & \multicolumn{3}{c}{AGN Magnitude}
&\colhead{}  
&\multicolumn{3}{c}{AGN Flux Density (mJy)} 
&\colhead{}
\\ \cline{3-5} \cline{7-9}
\colhead{ID}&\colhead{Object}
&\colhead{J}&\colhead{H}&\colhead{K}
&\colhead{}
&\colhead{J}&\colhead{H}&\colhead{K}
&\colhead{Host}
\\
\colhead{(1)} & \colhead{(2)} & \colhead{(3)} & \colhead{(4)}
& \colhead{(5)} & \colhead{} & \colhead{(6)} & \colhead{(7)} & 
\colhead{(8)} & \colhead{(9)} 
}
\startdata
2& PG 0052+251 	& 14.15& 13.34& 12.36
&&    3.49$\pm$0.06&    4.72$\pm$0.09&    7.58$\pm$0.14&   yes\\
3& PKS 0112$-$017 	& 16.93& 15.81& 15.66
&&    0.27$\pm$0.04&    0.49$\pm$0.06&    0.36$\pm$0.07&      \\
4& 3C 37 	& 16.73& 15.91& 15.10
&&    0.32$\pm$0.04&    0.44\phantom{$\pm$0.12}&    0.61$\pm$0.08&      \\
5& 3C 47 	& 16.35& 15.44& 14.55
&&    0.46$\pm$0.05&    0.68$\pm$0.07&    1.01$\pm$0.08&      \\
6& 4C 01.04 	& 15.84& 14.87& 13.82
&&    0.74$\pm$0.05&    1.15$\pm$0.09&    1.98$\pm$0.11&      \\
7& 4C 10.06 	& 15.35& 14.55& 13.80
&&    1.15$\pm$0.05&    1.55$\pm$0.07&    2.01$\pm$0.09&      \\
8& PKS 0403$-$13 	& 15.61& 14.82& 14.12
&&    0.91$\pm$0.05&    1.21$\pm$0.09&    1.50$\pm$0.08&      \\
9& 3C 110 	& 14.87& 14.73& 13.94
&&    1.80$\pm$0.07&    1.31$\pm$0.06&    1.77$\pm$0.08&      \\
\enddata
\tablecomments{
For objects without H-band host fraction, we take 2MASS profile fitting
magnitude as the AGN magnitude.
Values without uncertainties reflect the fact that the original
sources did not provide uncertainties.
}
\tablenotetext{a}{
Indicating whether we have host fraction from H-band observation.
}
\end{deluxetable}

\begin{deluxetable}{rlccc}
\tabletypesize{\scriptsize}
\tablecolumns{5}
\tablewidth{0pc}
\tablecaption{Far-UV \fuse\ Data
\label{tb:fuse}
}
\tablehead{
\colhead{}& \colhead{} 
&\multicolumn{2}{c}{Rest Wavelength (\AA)\tablenotemark{a}}
&\colhead{Flux\tablenotemark{b}}
\\ \cline{3-4}
\colhead{ID}&\colhead{Object}
&\colhead{$\lambda$1}&\colhead{$\lambda$2}
&\colhead{($10^{-14}$~\fluxl)}
\\
\colhead{(1)} & \colhead{(2)} & \colhead{(3)} & \colhead{(4)}
& \colhead{(5)} 
}
\startdata
 2 & PG 0052+251 &	   789  &  1028  &   1.67\\
13 & IRAS F07546+3928 &	   832  &  1083  &   0.72\\
15 & PG 0844+349 &	   856  &  1115  &   4.83\\
20 & PG 0947+396 &	   758  &   984  &   1.12\\
21 & PG 0953+414 &	   740  &   962  &   5.84\\
28 & PG 1100+772 &	   696  &   905  &   1.50\\
33 & PG 1116+215 &	   768  &  1010  &   6.58\\
36 & 3C 263 &	   549  &   721  &   1.19\\
42 & PG 1216+069 &	   678  &   892  &   1.16\\
43 & PG 1226+023 &	   787  &  1025  &  27.36\\
46 & PG 1259+593 &	   618  &   803  &   1.90\\
49 & PG 1322+659 &	   781  &  1015  &   1.21\\
51 & PG 1351+640 &	   837  &  1090  &   1.91\\
56 & PG 1402+261 &	   776  &  1020  &   3.15\\
58 & PG 1415+451 &	   811  &  1066  &   1.27\\
61 & PG 1440+356 &	   839  &  1103  &   5.27\\
64 & PG 1534+580 &	   885  &  1151  &   3.29\\
70 & PG 1626+554 &	   798  &  1050  &   0.67\\
73 & PG 1704+608 &	   664  &   864  &   0.16\\
74 & MRK 506 &	   874  &  1138  &   1.74\\
75 & 4C 34.47 &	   757  &   984  &   0.60\\
77 & MRK 509 &	   885  &  1145  &  11.01\\
80 & PG 2214+139 &	   855  &  1113  &   2.23\\
85 & PG 2349$-$014 &	   777  &  1011  &   1.84\\
\enddata
\tablenotetext{a}{Wavelength coverage in rest-frame.
}
\tablenotetext{b}{Mean flux density between 1050--1070\,\AA\ in
the observed-frame with an uncertain less than 5\%.  
There is not a common region in the rest-frame where all 
objects have fluxes from \fuse.
}
\end{deluxetable}

\begin{deluxetable}{rlllrlllrl}
\tabletypesize{\scriptsize}
\tablecolumns{10}
\tablewidth{0pc}
\tablecaption{X-ray Spectral Parameters (Single or Broken Power-law)
\label{tb:xray}
}
\tablehead{
\colhead{}& \colhead{} 
&\colhead{}  
&\colhead{}  
&\colhead{}  
&\colhead{}  
&\multicolumn{3}{c}{$f_\nu = f_0 E^\alpha$} 
&\colhead{}
\\ \cline{6-9}
\colhead{ID}&\colhead{Object}
&\colhead{E1}&\colhead{E2}&\colhead{E3}
&\colhead{$f_0$}&\colhead{$\alpha$}
&\colhead{$f_0$}&\colhead{$\alpha$}
&\colhead{Reference}
\\
\colhead{(1)} & \colhead{(2)} & \colhead{(3)} & \colhead{(4)}
& \colhead{(5)} 
& \colhead{(6)} & \colhead{(7)} & 
\colhead{(8)} & \colhead{(9)}  & \colhead{(10)}
}
\startdata
1& MC2 0042+101 &  0.1&     & 2.4
	& 1.09E-4&$ -1.48^{+0.97}_{-2.04}$
	&        &          
	& R,Br97\\
2& PG 0052+251 &  0.1&     & 2.4
	& 1.50E-3&$ -1.49^{+0.02}_{-0.04}$
	&        &          
	& R,Yu98\\
3& PKS 0112$-$017 &  0.1&     & 2.4
	& 6.06E-5&$ -0.88^{+0.56}_{-0.79}$
	&        &          
	& R,Br97\\
4& 3C 37 &  0.1&     & 2.4
	& 1.78E-4&$ -1.32^{+0.34}_{-0.34}$
	&        &          
	& R,Br97\\
5& 3C 47 &  2.0&     &10.0
	& 9.93E-5&$ -0.28$
	&        &          
	& C,Sh05\\
6& 4C 01.04 &  0.1&     & 2.4
	& 3.90E-4&$ -0.93^{+0.04}_{-0.04}$
	&        &          
	& R,Br97\\
7& 4C 10.06 &  0.1&     & 2.4
	& 9.08E-4&$ -1.13^{+0.52}_{-0.56}$
	&        &          
	& R,Br97\\
8& PKS 0403$-$13 &  0.1&     & 2.4
	& 4.03E-4&$ -0.60^{+0.06}_{-0.06}$
	&        &          
	& R,Br97\\
\enddata
\tablecomments{
E1, E2, and E3 are the 
observed-frame energies (in \kev) at which the power-law models are fit.
When broken power-law models are used, E2 and columns
(8)-(9) are needed to present them.
$f_0$ is flux density at 1\,\kev, in units of mJy ($10^{-26}$ \fluxhz);
$E$ in \kev.
Values without uncertainties reflect the fact that the original
sources did not provide uncertainties.
}
\tablerefs{
R,C,X indicate data sources, corresponding to 
ROSAT, Chandra, and XMM, respectively.
Br97---Brinkmann et al. 1997; Yu98---Yuan et al. 1998; La97---Laor et
al. 1997; Be06---Belsole et al. 2006; Ha96---Hardcastle et al. 2006; 
Sh05---Shi et al. 2005; Sa06---Sambruna et al. 2006;
Ga03---Gambill et al. 2003; Ta07---Tavecchio et al. 2007;
Si08---Siemiginowska et al. 2008; CSC---Chandra Source Catalog;
Pi05---Piconcelli et al. 2005; Po04---Porquet et al. 2004;
Fo06---Foschini et al. 2008
}
\end{deluxetable}


\begin{deluxetable}{crrrrrlrrr}
\tabletypesize{\scriptsize}
\tablecolumns{10}
\tablewidth{0pc}
\tablecaption{Parameters of Defining Bins in Each Waveband
\label{tb:bins}
}
\tablehead{
\colhead{Wavelength}
&\multicolumn{4}{c}{Radio-Loud}
&\colhead{}
&\multicolumn{4}{c}{Radio-Quiet}
\\ \cline{2-5} \cline{7-10}
\colhead{Band}
&\colhead{$\nu1$\tablenotemark{a}}
&\colhead{$\nu2$\tablenotemark{a}}
&\colhead{Bins\tablenotemark{b}}
&\colhead{$\Delta \nu$\tablenotemark{c}}
&\colhead{}
&\colhead{$\nu1$\tablenotemark{a}}
&\colhead{$\nu2$\tablenotemark{a}}
&\colhead{Bins\tablenotemark{b}}
&\colhead{$\Delta \nu$\tablenotemark{c}}
}
\startdata
radio  &  7.90 &  9.02 &   3 &  0.373	 &&  7.90 &  9.02 &   3 &  0.373\\
radio  &  9.18 &  9.55 &   1 &  0.370	 &&  9.13 &  9.51 &   1 &  0.380\\
radio  &  9.71 & 10.52 &   2 &  0.405	 &&  9.67 & 10.31 &   2 &  0.320\\
FIR    & 12.33 & 12.89 &   2 &  0.280	 && 12.28 & 12.81 &   2 &  0.265\\
MIR    & 13.06 & 13.90 &  30 &  0.028	 && 12.91 & 13.83 &  30 &  0.031\\
NIR    & 14.20 & 14.54 &   4 &  0.085	 && 14.15 & 14.49 &   3 &  0.113\\
UV/opt & 14.61 & 15.55 & 200 &  0.005	 && 14.55 & 15.55 & 200 &  0.005\\
X-ray  & 16.46 & 18.63 &   6 &  0.362	 && 16.88 & 18.41 &   4 &  0.383\\
\enddata
\tablecomments{
See \S\ref{sec:msed} on how the bins are defined in constructing composite
SEDs.
}
\tablenotetext{a}{Frequency ranges in Log(Hz).}
\tablenotetext{b}{Number of bins in the range.}
\tablenotetext{c}{Bin size in Log(Hz), $\Delta\nu = (\nu2-\nu1)/Bins$.}
\end{deluxetable}


\end{document}